\documentclass[11pt]{article}
\usepackage{amsmath}
\usepackage{amsfonts}

\usepackage{amssymb,graphicx}

\usepackage{amssymb}

\newtheorem{prop}{Proposition}
\newtheorem{thm}{Theorem}

\newcommand{\bbr}{{\mathbb R}}

\newcounter{mnotecount}[section]

\renewcommand{\themnotecount}{\thesection.\arabic{mnotecount}}

\newcommand{\mnote}[1]
{\protect{\stepcounter{mnotecount}}$^{\mbox{\footnotesize
$
\bullet$\themnotecount}}$ \marginpar{
\raggedright\tiny\em
$\!\!\!\!\!\!\,\bullet$\themnotecount: #1} }

\def\p{\partial}
\def\tg{\tilde{g}}
\def\ta{\tilde{a}}
\def\tp{\tilde{p}}
\def\tb{\tilde{b}}
\def\tG{\tilde{\Gamma}}
\def\tR{\tilde{R}}

\def\tk{\tilde{k}}
\def\tnabla{\widetilde{\nabla}}
\def\tT{\tilde{T}}
\def\trho{\tilde{\rho}}
\def\tu{\tilde{u}}
\def\tC{\tilde{C}}
\def\th{\tilde{h}}
\def\tomega{\tilde{\omega}}
\def\tk{\tilde{k}}
\def\tsigma{\tilde{\sigma}}
\def\be{\begin{equation}}
\def\ee{\end{equation}}
\def\bea{\begin{eqnarray}}
\def\eea{\end{eqnarray}}
\def\bean{\begin{eqnarray*}}
\def\eean{\end{eqnarray*}}
\def\hZ{\hat Z}
 \DeclareMathOperator{\diag}{diag}
\usepackage{authblk}
\usepackage{xcolor} 

\begin{document}

\title{Homogeneous solutions to the Einstein-matter equations with a magnetic field and a conformal gauge singularity}

\author[1]{Ho Lee\footnote{holee@khu.ac.kr}}
\author[2]{Ernesto Nungesser\footnote{em.nungesser@upm.es}}
\author[3]{John Stalker\footnote{stalker@maths.tcd.ie}}
\author[4]{Paul Tod\footnote{tod@maths.ox.ac.uk}}

\affil[1]{Department of Mathematics and Research Institute for Basic Science, College of Sciences, Kyung Hee University, Seoul 02447, Republic of Korea}
\affil[2]{M2ASAI, Universidad Polit{\'e}cnica de Madrid, ETSI Navales, Avda. de la Memoria
4, 28040 Madrid, Spain}
\affil[3]{School of Mathematics, Trinity College, Dublin 2, Ireland and Hamilton Mathematics Institute, Dublin 2, Ireland}
\affil[4]{Mathematical Institute, University of Oxford, Oxford OX2 6GG}

\maketitle

\begin{abstract}
We study massless solutions to the Einstein equations coupled to different matter models with a magnetic field and a conformal gauge singularity assuming spatial homogeneity with three commuting spatial translations. We show that there are no solutions in the case that the matter model is a radiation fluid. If the matter is described via kinetic theory we obtain that there exist unique solutions to the Einstein-Vlasov system and the Einstein-Boltzmann system for a certain range of soft potentials. For both the Vlasov and the Boltzmann case we also obtain asymptotic expansions close to the initial conformal gauge singularity.
\end{abstract}

\section{Introduction}
There is a general consensus that General Relativistic cosmological models, to be realistic models of the actual universe, must have initial curvature singularities at which the metric and some at least of the matter variables must be singular. The most general singularities of a Lorentzian manifold subject to the Einstein field equations can apparently be extremely complicated, with oscillations of the geometric quantities showing chaotic behaviour in space and time on the approach to the singularity. However, there are physical arguments that the initial singularity of the actual universe does not have this character – it was much less complicated than it is mathematically allowed to be (see \cite{P1} for these arguments). 
Penrose \cite{P1} has conjectured that this constraint on the initial singularity could be expressed as finiteness of the Weyl curvature tensor at the initial singularity, while the Ricci tensor, which is directly related to the matter content in General Relativity, is singular there. This conjecture can be conveniently expressed by the assumption that the initial singularity is a \emph{conformal gauge singularity} \cite{LT} by which is meant that there is a conformal rescaling of the physical metric, which is becoming singular, to obtain an unphysical metric which is nonsingular, and therefore defined on a larger manifold, giving an extension through the singularity and adding a boundary to the space-time. Thus the presence of the singularity is attributed to the choice of the physical metric within its conformal class, the class of conformally-related metrics – the singularity is due to a choice of conformal gauge. This notion then raises the attractive possibility of posing an initial value problem with regular initial data at the initial boundary which has been added.
This program has been successfully carried through, with the proof of well-posedness for the Cauchy problem with data at the initial boundary of an unphysical metric, and with the physical metric singular on approach to the boundary, for a variety of matter models taken as sources for the Einstein equations. Thus polytropic perfect fluids were considered in \cite{AT1}, the massless Einstein-Vlasov equations (i.e. massless collisionless kinetic theory) in \cite{AT2} and \cite{A} and with a cosmological constant added in \cite{T3}. In  \cite{LNT1} and \cite{LNST} the Vlasov equation was replaced by the Boltzmann equation, so that the matter model is massless kinetic theory but now with collisions, and well-posedness was proved for the Cauchy problem with a certain class of scattering kernels and in certain very symmetric space-times: spatially homogeneous and isotropic in \cite{LNT1} and spatially homogeneous with three commuting spatial translations in \cite{LNST} (this latter symmetry is known in the literature of General Relativity as \emph{Bianchi type I}). 
The reason for restricting to \emph{massless} kinetic theory comes from a physical consideration, that near the initial singularity the particles implicit in the kinetic theory will have energies very large compared to their rest-mass, which can therefore be neglected. The disadvantage of this assumption is that the standard existence theorems for the Boltzmann and Einstein-Boltzmann equations (e.g. \cite{Bancel} and \cite{BC}) assume massive particles, and there don’t seem to be existence theorems of similar generality for massless particles in the literature. With rest-mass zero, there is an extra singularity at zero energy in the Boltzmann equation which has to be dealt with. This proved possible with the scattering kernels considered in \cite{LNT1} and \cite{LNST} and so we retain those in this work.
There is a large literature on cosmological singularities, with different matter models and different symmetry assumptions. One area that might be thought to be close to our program is that of \emph{quiescent cosmology} \cite{AR, Barrow, FRS, Ring2}. Here the initial singularity is constrained in a different way: the assumption is that oscillatory behaviour is absent, but there is not necessarily a conformal rescaling of the physical metric that would make it regular: while there is no oscillation, the different components of the metric may become singular (or zero) at different rates. It is often still possible to give data `at the singularity’ \cite{FRS, Ring2} but, unlike with a conformal gauge singularity, the data doesn’t have the character of, for example, an initial 3-metric at an initial boundary.

This article is an extension of \cite{LNST} to other matter models. We work with the same symmetry class, along with a homogeneous magnetic field satisfying the source-free Maxwell equations, and either a radiation fluid (that is, a perfect fluid with pressure equal to one-third of density) or a kinetic theory source specified by a distribution function $f$ and either collisionless (the Vlasov case) or with collisions and subject to the Boltzmann equation. This symmetry class fits very well with a homogeneous magnetic field and this combined with other fields has been much studied \cite{Ayissi, HU, Jacobs, LB, LRT}, though not, as here, with an initial conformal gauge singularity. 
Our method leans heavily on the existence and uniqueness theorem for Fuchsian systems, Theorem 4 of \cite{LNST}, which is a strengthening of the earlier result of \cite{RS}. In the different cases, we first conformally rescale the metric and matter fields and then choose variables so as to cast the Einstein equations into Fuchsian form and, in the relevant case, to regularise the Boltzmann equation. The Fuchsian system imposes certain conditions, conveniently also called Fuchsian conditions, on the initial data so it is necessary to show that there exist initial data satisfying these conditions. Then Theorem 4 of \cite{LNST} requires differentiability of the coefficients of the Fuchsian system, which must be verified, and requires conditions on the eigenvalues of a matrix defined by the Fuchsian system, one condition sufficient for the existence of a solution bounded at time zero, i.e. near the (removed) singularity, and a stronger one sufficient for existence of a solution differentiable at time zero.
In the case of the radiation fluid and magnetic field, with or without cosmological constant, the Fuchsian conditions are seen to be so strong that there do not exist initial data with a nonzero magnetic field. After observing this, therefore, we drop the perfect fluid case and consider the two kinetic theory cases. Both of these have a satisfactory solution: given any distribution function compactly supported in momentum space away from the origin we obtain initial data to the rescaled Einstein-Vlasov and Einstein-Boltzmann system. The precise statement can be found in Proposition 1 and Proposition 2 respectively.  Based on this we prove the existence and uniqueness of solutions to the rescaled Einstein-Vlasov and Einstein-Boltzmann system. If the magnetic field is small in a certain sense which will be specified later, we even obtain that the solutions to the Einstein equations are differentiable in the rescaled, i.e. unphysical, coordinates. The precise statements are Theorem 1 and Theorem 3 respectively. This implies in particular that in the absence of a magnetic field we obtain a stronger result in the Vlasov case than was obtained previously in \cite{AT2}. The reason for this is that we apply Theorem 4 of \cite{LNST} which is stronger than the theorem used in \cite{AT2} which is based on \cite{RS}.

The results concerning the rescaled Einstein-Vlasov and rescaled Einstein-Boltzmann system are also expressed in terms of the unrescaled, physical quantities. In particular we obtain for both cases asymptotic expansions close to the initial conformal gauge singularity which can be found in Theorem 2 and Theorem 4 respectively. As a result of the differentiability of the solutions for small magnetic fields we obtain sharper asymptotics in that case. Note that we have allowed in all cases the existence of a cosmological constant, which however does not play an important role in the arguments.

In conclusion, we have shown that \cite{AT1} cannot be generalised to include a magnetic field in the case of a radiation fluid and we have generalised \cite{AT2, LNST} to include a magnetic field. The main contribution of this paper is the generalisation of the Einstein-Vlasov case \cite{AT2}, since once achieved, the generalisation of the Einstein-Boltzmann case follows straighforwardly from the results of \cite{LNST}. Note that in both generalisations we observe the same remarkable feature which happens also in absence of a magnetic field: the initial distribution function determines uniquely the initial conformal metric.

The paper is organised as follows. In the next section we introduce all the relevant equations where we couple the Einstein equations with a radiation fluid, collisionless matter and with matter which interacts via the Boltzmann equation for a certain range of soft potentials. In Section 3 introducing certain time variable changes we obtain the conformally rescaled equations for all the cases coming to the conclusion that in the presence of a magnetic field there is no solution to the Einstein equations coupled to a radiation fluid and a conformal gauge singularity.  In the end of that section we will state our main theorems concerning solutions to the Einstein-Vlasov and the Einstein-Boltzmann system which will be proven in Section 4 and 5 respectively. In Section 4 we consider the Vlasov case and prove the main results for that case showing that all conditions to apply Theorem 4 of \cite{LNST} are satisfied. Using the results concerning the Boltzmann equation of \cite{LNST} and the results developed in Section 4 we prove the main results concerning the Boltzmann case in the last section.

\medskip

\section{The equations}
\subsection{The metric and curvature}
We follow our previous conventions. The Bianchi I physical metric is
\be\label{1}
\tg=-dt^2+\ta_{ij}dx^idx^j,
\ee
with $i,j,\ldots= 1,2,3$ and the $t$-coordinate labelled 0, and where $\ta_{ij}$ is a function only of $x^0=t$. The metric is tilded so that the rescaled metric, with which we mainly work, can be untilded. Set $\tb^{ij}$ inverse to $\ta_{ij}$ so that
\be\label{2}
\tb^{ik}\ta_{kj}=\delta^i_{\,j}.
\ee
Introduce the rate of change, which is a constant multiple of the second fundamental form, and its trace as
\be\label{3}
\tk_{ij}:=\frac{d}{dt}\ta_{ij},\;\;\tk:=\tb^{ij}\tk_{ij}.
\ee
From 
\be
{\Gamma^{\alpha}}_{\beta \gamma} = \frac12 g^{\alpha\delta} (g_{\delta \beta, \gamma} + g_{\delta \gamma, \beta} - g_{\beta \gamma, \delta}),
\ee
where Greek indices run from 0 to 3 we obtain that the only nonzero Christoffel symbols are, 
\be\label{4}
 {{\tG^i}}_{\ 0j}= {\tG^i}_{\ j0} =\frac12\tb^{il}\tk_{lj},\;\;\;{\tG^0}_{\ ij}=\frac12\tk_{ij}.
\ee
Now the components of the Ricci tensor are [(3.4.5) of \cite{Wald}] 
\begin{align}
R_{\mu \rho} = {\Gamma^{\nu}}_{\mu \rho, \nu} - {\Gamma^{\nu}}_{\nu \rho, \mu} + {\Gamma^{\alpha}}_{\mu \rho} {\Gamma^{\nu}}_{\alpha \nu} - {\Gamma^{\alpha}}_{\nu\rho} {\Gamma^{\nu}}_{\alpha \mu}.
\end{align}
In our case we obtain 
\bea\label{5}
\tR_{00}&=&-\frac12\frac{d\tk}{dt}-\frac14\tb^{im}\tb^{jn}\tk_{ij}\tk_{mn},\\\label{6}
\tR_{0i}&=&0,\\
\label{7} \tR_{ij}&=&\frac12\frac{d}{dt}\tk_{ij}+\frac14\tk\tk_{ij}-\frac12\tk_{im}\tb^{mn}\tk_{nj}.
\eea

From \eqref{5}--\eqref{7} we can compute $\tilde{R}$:
\begin{align}
\nonumber \tilde{R} &= \tilde{R}_{\alpha\beta} \tilde{g}^{\alpha \beta} = - \tilde{R}_{00} + \tilde{b}^{ij} \tilde{R}_{ij} \\
\label{R}  &= \frac12\frac{d\tk}{dt}-\frac14\tb^{im}\tb^{jn}\tk_{ij}\tk_{mn}+ \frac12 \tilde{b}^{ij}\frac{d}{dt}\tk_{ij}+\frac14\tk^2.
\end{align}
Now since 
\begin{align}
\frac{d\tilde{k} }{dt} = \frac{d}{dt} \left ( \tilde{b}^{ij} \tilde{k}_{ij} \right) = - \tilde{b}^{mi}\tilde{b}^{nj}\tilde{k}_{mn} \tilde{k}_{ij} + \tilde{b}^{ij} \frac{d}{dt} \tilde{k}_{ij},
\end{align}
we can express \eqref{R} as
\begin{align}
 \tilde{R}  = \frac{d\tk}{dt} + \frac14 \tilde{k}^2 + \frac14 \tilde{b}^{im}\tilde{b}^{jn} \tilde{k}_{mn} \tilde{k}_{ij}.
\end{align}
This implies
\begin{align}\label{einstein00}
\tilde{G}_{00} = \tilde{R}_{00} - \frac12 \tilde{g}_{00} \tilde{R} =  \frac18 \tilde{k}^2 - \frac18  \tilde{b}^{im}\tilde{b}^{jn} \tilde{k}_{mn} \tilde{k}_{ij}.
\end{align}

\medskip

\subsection{The various sources}
We'll be interested in three kinds of source, all of which have a trace-free stress tensor and we will also allow a cosmological constant.

\medskip

\subsubsection{The radiation perfect fluid}
The first kind of source, which is instructive but turns out to have no interesting solutions, is the radiation perfect fluid, for which the stress tensor is
\be\label{8a}
\tT^{\mbox{rad}}_{\alpha\beta}=\frac13\trho(4\tu_\alpha \tu_\beta+\tg_{\alpha\beta}),
\ee
where $\tu^\alpha$ is the (unit, time-like, future pointing) velocity vector, which we shall always assume parallel to the coordinate vector field $\p/\p t$, and $\trho$ is the energy density, assumed non-negative. 
Since this $\tu^\alpha$ is necessarily geodesic (from (\ref{1})), the only consequence of the conservation equation for this stress tensor is
\be\label{8b}
\trho \ta ^{2/3}=\trho_0,
\ee
where $\ta=\mbox{det}(\ta_{ij})$ and $\trho_0$ is a constant of integration. Decomposing in the coordinate basis, we note that:
\bea
\label{rad1}\tT^{\mbox{rad}}_{00}&=&\trho,\\
\label{rad2}\tT^{\mbox{rad}}_{0i}&=&0,\\
\label{rad3}\tT^{\mbox{rad}}_{ij}&=&\frac{1}{3}\trho\ta_{ij}.
\eea

\medskip

\subsubsection{The source-free magnetic field}

The second kind of source is source-free magnetic field. In the coordinate basis, a purely magnetic, homogeneous Maxwell field is represented by a two-form
\be\label{9}
{\bf F}=F_{ij}dx^i\wedge dx^j,
\ee
where the components $F_{ij}=-F_{ji}$ are independent of $x^i$ to preserve the spatial homogeneity. To satisfy the first set of Maxwell equations, we need ${\bf F}$ to be closed which requires the $F_{ij}$ to be independent of $t$, whence they are constant. The other set of Maxwell equations, with sources set to zero, namely
\begin{align}
\tnabla^\alpha F_{\alpha\beta}=0,
\end{align}
are now automatic, given the Christoffel symbols (\ref{4}).
It will be useful to write the constants $F_{ij}$ in terms of a vector $h^i$ of constants which we can think of as the magnetic field vector by
\be\label{9a}
F_{ij}=\epsilon_{ijk}h^k,\ee
where $\epsilon_{ijk}$ is the alternating tensor with $\epsilon_{123}=1$. 
Note that this is not the volume form for the spatial metric, which is
\begin{align}
\tilde{\eta}_{ijk}=\ta^{1/2}\epsilon_{ijk},
\end{align}
and is therefore time-dependent. It's useful to note the following identities:
\begin{align}
\tb^{jq}\epsilon_{ijm}\epsilon_{pqn}=\frac{1}{\ta}(\ta_{ip}\ta_{mn}-\ta_{in}\ta_{mp}),\quad \tb^{ip}\tb^{jq}\epsilon_{ijm}\epsilon_{pqn}=\frac{2}{\ta}\ta_{mn}.
\end{align}

The stress tensor for the Maxwell field is 
\be\label{10}
\tT^{\mbox{max}}_{\alpha\beta}=\frac{1}{2\pi}(\tg^{\gamma\delta}F_{\alpha\delta}F_{\beta\gamma}
-\frac14\tg_{\alpha\beta}\tg^{\gamma\delta}\tg^{\lambda\mu}F_{\gamma\lambda}F_{\delta\mu}).
\ee
The components can be written in terms of the magnetic field vector as
\bea
\tT^{\mbox{max}}_{00}&=&\frac{1}{4\pi\ta}\ta_{ij}h^ih^j,\\
\tT^{\mbox{max}}_{0i}&=&0,\\
\tT^{\mbox{max}}_{ij}&=&\frac{1}{4\pi\ta}(\ta_{ij}\ta_{mn}-2\ta_{im}\ta_{jn})h^mh^n,
\eea
and the conservation equation gives nothing new.

\medskip

\subsubsection{The kinetic part}

The third source is matter satisfying the assumptions of kinetic theory. With conventions as in \cite{LNST} and a distribution function $f=f(t, p_i)$ which is independent of $x^i$, to be consistent with the Bianchi I symmetry. The distribution function on the cotangent bundle is supported on the null-cone bundle with fibre $N_x$. Thus the stress tensor is
\be\label{11}
\tT^{\mbox{kt}}_{\alpha\beta}=\int_{N_x}fp_\alpha p_\beta\tomega_p,\ee
where $\tomega_p$ is the invariant volume measure given explicitly below \eqref{vol} and the distribution function satisfies the Boltzmann equation
\be\label{12}
\mathcal{L}_{\tg}(f)=\tC(f,f),
\ee
see \cite{LNST} for details of $\tC$. In particular we shall here consider the same set of scattering cross-sections as in \cite{LNST}, namely 
\be\label{13}
\tsigma=\th^{-\gamma},
\ee
where $\th$ is the physical relative momentum, which is 
\begin{align}
\th = \sqrt{-2\tg^{\alpha\beta}p_\alpha q_\beta},
\end{align}
 for a collision between massless particles of momenta $p_\alpha,q_\beta$, and $\gamma$ is a real number in the range $1<\gamma<2$. It's worth noting here that for Bianchi I metrics there is a great simplification in the Liouville operator $\mathcal{L}$ which becomes just 
\begin{align}
{\mathcal{L}}_{\tg} f=\tp^0\frac{\partial f}{\partial t},
\end{align}
where 
\begin{align}\label{p0}
\tp^0=(\tb^{mn}p_mp_n)^{1/2}.
\end{align}
In particular in the collisionless or Vlasov limit the distribution function $f$ becomes constant in time.
By expanding 
\begin{align}\label{vol}
\tomega_p=\frac{1}{p^{0}\sqrt{\tilde{a}}}\, d^{3}p = \frac{1}{p^{0}\sqrt{\tilde{a}}}\, dp_1 dp_2 dp_3,
\end{align}
 and substituting for $\tp^0$ with \eqref{p0} we can write the components of the kinetic theory stress tensor as 
\bea
\tT^{\mbox{kt}}_{00}&=&\frac{1}{\sqrt{\ta}}\int f (\tb^{mn}p_mp_n)^{1/2}d^3p,  \\
\tT^{\mbox{kt}}_{0i}&=&- \frac{1}{\sqrt{\ta}}\int fp_id^3p,   \\
\tT^{\mbox{kt}}_{ij}&=&\frac{1}{\sqrt{\ta}}\int f p_ip_j(\tb^{mn}p_mp_n)^{-1/2}d^3p,
\eea
and again, the conservation equation gives nothing new. However it is worth observing that with or without collisions, the integral
\begin{align}
\int fp_id^3p,
\end{align}
is a constant of the motion and so will vanish at all times if it vanishes at one, and we need this to be zero for satisfaction of the $(0i)$-Einstein equation.

\medskip

\section{New time coordinates and conformal rescaling}
\subsection{Behaviour of the metric and curvature}
We do this in two stages. Since the stress tensors under consideration are all trace-free, we can start with the redefinition of the time-coordinate, familiar from \cite{LNST}: 
\begin{align} \label{ttau}
\tau=(2t)^{1/2},
\end{align}
and the conformal rescaling
\be\label{13a}
a_{ij}=\tau^{-2}\ta_{ij},\;\;b^{ij}=\tau^2\tb^{ij}.
\ee
We explore the consequences of this first, and then later modify the time-coordinate again to regularise the Boltzmann equation.
We redefine the rate of change tensor as
\begin{align}
k_{ij}:=\frac{d}{d\tau}a_{ij}=\frac{dt}{d\tau}\frac{d}{dt} \left(\frac{1}{2t}\ta_{ij}\right)=\frac{1}{\tau}(\tk_{ij}-2a_{ij}),
\end{align}
or
\be\label{14}
\tk_{ij}=\tau k_{ij}+2a_{ij}.\ee
From the trace of this, with 
\begin{align}
\tk=\tb^{ij}\tk_{ij}, \quad k=b^{ij}k_{ij}, 
\end{align}
we obtain
\be\label{15}
\tk=\frac{1}{\tau}k+\frac{6}{\tau^2}.
\ee
From \eqref{5}--\eqref{7} the physical Ricci tensor can be written
\bea
\label{15b}
\tR_{00}&=& -\frac{1}{2\tau^2}\frac{dk}{d\tau}-\frac{1}{4\tau^2}b^{ip}b^{jq}k_{ij}k_{pq}-\frac{k}{2\tau^3}+\frac{3}{\tau^4},\\
\label{16} 
\tR_{0i}&=&0,\\ 
\tR_{ij}&=&\frac12\frac{dk_{ij}}{d\tau}+\left(\frac{1}{\tau}+\frac14k\right)k_{ij}
-\frac12k_{im}b^{mn}k_{nj}+\left(\frac{1}{\tau^2}+\frac{k}{2\tau}\right)a_{ij}\label{17}.
\eea

\medskip

\subsection{Behaviour of the source-terms}
These are all much the same but worth collecting together. For the radiation stress tensor we have (\ref{8b}) so that
\begin{align}
\trho=\tau^{-4} a ^{-2/3}\trho_0,
\end{align}
with 
\begin{align}\label{det}
a=\mbox{det}(a_{ij})=\tau^{-6}\ta,
\end{align}
 and then \eqref{rad1}-\eqref{rad3} turns into
\bea
\tT^{\mbox{rad}}_{00}&=&   \tau^{-4} a^{-2/3}\trho_0,\\
\tT^{\mbox{rad}}_{0i}&=&0,\\
\tT^{\mbox{rad}}_{ij}&=&\frac{1}{3}\tau^{-2} a^{-2/3}\trho_0a_{ij}.
\eea
For the Maxwell stress tensor, $h^i$ is a constant vector and unchanged by conformal rescaling, so that
\bea
\tT^{\mbox{max}}_{00}&=&\tau^{-4}\frac{1}{4\pi a}a_{ij}h^ih^j,\\
\tT^{\mbox{max}}_{0i}&=&0,\\
\tT^{\mbox{max}}_{ij}&=&\tau^{-2}\frac{1}{4\pi a}(a_{ij}a_{mn}-2a_{im}a_{jn})h^mh^n.
\eea
Finally, for the kinetic theory stress tensor, we find
\bea\label{13b}
\tT^{\mbox{kt}}_{00}&=&\tau^{-4}\frac{1}{\sqrt{a}}\int f (b^{mn}p_mp_n)^{1/2}d^3p,\\
\tT^{\mbox{kt}}_{0i}&=&-\tau^{-3}\frac{1}{\sqrt{a}}\int f p_id^3p , \\\label{13c}
\tT^{\mbox{kt}}_{ij}&=&\tau^{-2}\frac{1}{\sqrt{a}}\int f p_ip_j(b^{mn}p_mp_n)^{-1/2}d^3p\label{13d}.
\eea

\medskip

\subsection{The rescaled Einstein equations}

Since the sum of the stress tensors is trace-free, we obtain that the physical Ricci scalar is $\tilde{R}=4\Lambda$ and we obtain the Einstein equations by equating the physical Ricci tensor to $8\pi$ times the sum of the chosen physical stress tensors (setting the Newton constant $G$ to one) and $\Lambda \tilde{g}_{\alpha\beta}$, i.e. 
\begin{align}
\tilde{R}_{\alpha\beta} = 8\pi \tilde{T}_{\alpha\beta} + \Lambda \tg_{\alpha \beta}. 
\end{align}

 For the proof of existence we need this system to be Fuchsian but it isn't yet automatically so. To see this, consider first the $(ij)$-equation. This has a second-order pole in $\tau$ which wouldn't be compatible with a Fuchsian system, but it can be written as a first order pole
 \begin{align}
 \frac{2}{\tau}Z_{ij}, \quad Z_{ij} = \tau 8\pi \tilde{T}_{ij} - \frac{1}{\tau} a_{ij},
 \end{align}
 so
\be\label{19}Z_{ij}=\frac{1}{\tau}\left[\frac{8\pi}{\sqrt{a}}\int f p_ip_j\frac{d^3p}{(b^{mn}p_mp_n)^{1/2}}
+\frac{2}{a}(a_{ij}a_{mn}-2a_{im}a_{jn})h^mh^n-a_{ij}\right],\ee
when the source is magnetic field plus kinetic theory, or

\be\label{20}
Z_{ij}=\frac{1}{\tau}\left[\left(\frac{8\pi\trho_0}{3a^{2/3}}+\frac{2}{a}a_{pq}h^ph^q-1\right)a_{ij}-\frac{4}{a}a_{ip}a_{jq}h^ph^q\right],\ee
when the source is magnetic field plus radiation fluid.

The method in \cite{AT2} and \cite{LNST} was to add $Z_{ij}$ to the list of variables and obtain its evolution equation, whereupon the whole system of Einstein equations becomes a Fuchsian system. This in particular requires the tensor in the numerator of $Z_{ij}$ to vanish at $\tau=0$ and this becomes one of the Fuchsian conditions on the data. We shall see that this method continues to work with kinetic theory and a magnetic field. With a radiation fluid and a magnetic field we will see that there are no solutions with a conformal gauge singularity as 
follows from (\ref{20}): set the numerator to zero, then we require
\begin{align}
\left(\frac{8\pi\trho_0}{3a^{2/3}}+\frac{2}{a}a_{pq}h^ph^q-1\right)a_{ij}=\frac{4}{a}a_{ip}a_{jq}h^ph^q,
\end{align}
at $\tau=0$. Since the right-hand-side, if nonzero, is a matrix of rank one then necessarily the metric initially is degenerate, which we can't allow. If we set $h^i$ zero then there are solutions with both sides zero, so that $\trho_0$ is fixed by $a$, and these were found in \cite{AT1} but they're not of interest in this article as the magnetic field is zero. Henceforth we'll drop the radiation case.
Now we need the evolution equation for $Z_{ij}$. Having in mind that 
\begin{align}
\frac{d}{d\tau} a= k a, \quad  \frac{d}{d\tau}b^{ij}= - {b}^{im}{b}^{jn} {k}_{mn} ,
\end{align}
we obtain the following evolution equation for $Z_{ij}$:
\begin{align}\nonumber\frac{d Z_{ij}}{d\tau}&=-\frac{1}{\tau}Z_{ij}-\frac{1}{\tau}k_{ij}+\frac{1}{\tau}k_{pq}\frac{4\pi}{\sqrt{a}}b^{mp}b^{nq}\int \frac{fp_ip_jp_mp_n}{(b^{rs}p_rp_s)^{3/2}}d^3p\\\nonumber
&-\frac{1}{\tau}k\frac{4\pi}{\sqrt{a}}\int\frac{fp_ip_j}{(b^{rs}p_rp_s)^{1/2}}d^3p+\frac{1}{\tau}\frac{8\pi}{\sqrt{a}}\int \frac{\partial f}{\partial\tau}\frac{p_ip_j}{(b^{mn}p_mp_n)^{1/2}}d^3p\\
\label{22}&+\frac{2}{a\tau} \left[ a_{ij}k_{mn}+a_{mn}k_{ij}-2a_{im}k_{jn}-2a_{jn}k_{im} +k(2a_{im}a_{jn}-a_{ij}a_{mn})\right] h^mh^n.
\end{align}

Then the Einstein equations are 

\bea
\frac{dk}{d\tau}&=&-\frac{k}{\tau}-\frac{2b^{ij}Z_{ij}}{\tau}-\frac12b^{im}k_{ij}b^{jn}k_{mn} +2 \tau^2 \Lambda,\\
\int f p_id^3p&=&0,\\
\frac{dk_{ij}}{d\tau}&=&-\left(\frac{2}{\tau}+\frac{k}{2}\right)k_{ij}+\frac{2}{\tau}Z_{ij}-\frac{1}{\tau}b^{mn}k_{mn}a_{ij}+k_{im}b^{mn}k_{nj}+2\Lambda \tau^2 a_{ij}.
\eea

We can solve the Einstein-Vlasov case now, since in the absence of collisions the $\partial f/\partial\tau$-term in (\ref{22}) vanishes and we have a closed system, with only first-order poles in $\tau$. 

\medskip

\subsection{The Einstein constraint equations}

The momentum constraint equation is
\begin{align*} 
\tilde{G}_{0i} - 8 \pi \tilde{T}_{0i} = 0,
\end{align*}
which is equivalent to
\begin{align} \label{mconstraintphys}
  \frac{1}{\sqrt{\tilde{a}}}\int fp_id^3p = 0,
\end{align}
or in terms of $\tau$
\begin{align} \label{mconstraint}
\frac{1}{\tau^{3} \sqrt{{a}}}\int fp_id^3p = 0.
\end{align}
Thus we impose $\int fp_id^3p = 0$ as a condition on the initial $f$ and this continues to hold since the Boltzmann equation satisfies
\begin{align}\label{emconservation}
 \int  p_{\alpha} \frac{\partial f}{\partial t} d^3p = 0,
 \end{align}
  which expresses energy-momentum conservation in collisions. On the other hand we have the Hamiltonian constraint as
\begin{align}
\nonumber& \tilde{H} = \tilde{G}_{00}-\Lambda - 8 \pi \tilde{T}_{00} = 0 ,\\
\label{tildeH}  \iff &  \tilde{H} = \frac18 \tilde{k}^2 - \frac18  \tilde{b}^{im}\tilde{b}^{jn} \tilde{k}_{mn}  \tilde{k}_{ij} -\Lambda-  \frac{2}{\ta}\ta_{ij}h^ih^j- \frac{8\pi}{\sqrt{\ta}}\int f (\tb^{mn}p_mp_n)^{1/2}d^3p  =0.
\end{align}
One can compute using \eqref{5} and \eqref{7} that
\begin{align}
\nonumber\frac{d}{dt} \tilde{H} =& \frac14 \tk \frac{d \tk}{dt}  + \frac14    \tb^{jn}\tb^{ip}\tb^{mq}\tk_{pq} \tk_{mn}\tk_{ij}  - \frac14 \tilde{b}^{im}\tilde{b}^{jn} \tilde{k}_{mn} \frac{d}{dt}  \tk_{ij}+ 8 \pi \tk  \tT^{\mbox{max}}_{00} -  \frac{2}{\ta}\tk_{ij}h^ih^j \\
\nonumber& + 4\pi \tk \tilde{T}_{00}^{\mbox{kt}}  + 4\pi \tilde{b}^{mi}\tilde{b}^{nj} \tk_{ij} \tilde{T}_{mn}^{\mbox{kt}} \\
\nonumber= &\frac12 \tk \left (-\frac14\tilde{b}^{im}\tilde{b}^{jn}\tk_{ij}\tk_{mn}-\tR_{00} \right)  + \frac12 \tilde{b}^{ip}\tilde{b}^{jq}\tk_{pq} \left( \frac14\tk\tk_{ij}-\frac12\tk_{im}\tb^{mn}\tk_{nj} -  \tR_{ij}\right)\\
\nonumber&+ \frac14 \tb^{jn}\tb^{ip}\tb^{mq}\tk_{pq} \tk_{mn}\tk_{ij} + 8 \pi \tk  \tT^{\mbox{max}}_{00} -  \frac{2}{\ta}\tk_{ij}h^ih^j + 4\pi \tk \tilde{T}_{00}^{\mbox{kt}}  + 4\pi \tilde{b}^{mi}\tilde{b}^{nj} \tk_{ij} \tilde{T}_{mn}^{\mbox{kt}}\\
\nonumber= & -  \frac12 \tk \tR_{00} - \frac12 \tk^{ij}  \tR_{ij} + 8 \pi \tk  \tT^{\mbox{max}}_{00} -  \frac{2}{\ta}\tk_{ij}h^ih^j + 4\pi \tk \tilde{T}_{00}^{\mbox{kt}}  + 4\pi \tilde{b}^{mi}\tilde{b}^{nj} \tk_{ij}  \tilde{T}_{mn}^{\mbox{kt}}\\
=&-\frac12 \tilde{k} \tilde{H},
\end{align}
where we again used the energy-momentum conservation of collisions \eqref{emconservation} to remove the term with $\frac{\partial f}{\partial t}$. We see that if the Hamiltonian constraint equation is satisfied initially it will always be satisfied.

Now using \eqref{13a}, \eqref{14}, \eqref{15} and \eqref{det} equation \eqref{tildeH} turns into

\begin{align}
\nonumber H= \frac18 \left(\frac{k}{\tau} + \frac{6}{\tau^2}\right)^2 - \frac{1}{8\tau^{4}}  (\tau k_{ij}+2a_{ij}) (\tau b^{mi}b^{nj}k_{mn}+2b^{ij})
-\Lambda - \frac{2}{ a \tau^4}a_{ij}h^ih^j \\
- \frac{8\pi}{\tau^4 \sqrt{a}}\int f (b^{mn}p_mp_n)^{1/2}d^3p  =0,
\end{align}
which can be simplified to
\begin{align}
\tau^4 H = -\Lambda \tau^4 + \frac18 \tau^2 \left( k^2  - k_{ij}b^{mi}b^{nj}k_{mn}\right)  + k \tau  +3 -  \frac{2}{a}a_{ij}h^ih^j
- \frac{8\pi}{\sqrt{a}}\int f (b^{mn}p_mp_n)^{1/2}d^3p =0.
\end{align}
This means that for $\tau=0$ we have
\begin{align}\label{Hconstraint}
0= 3 -  \frac{2}{a}a_{ij}h^ih^j- \frac{8\pi}{\sqrt{a}}\int f (b^{mn}p_mp_n)^{1/2}d^3p.
\end{align}

\medskip

\subsection{Main results}

Now that all relevant equations have been introduced and denoting by $S_2(\bbr^3)$ the space of $ 3 \times 3 $ symmetric matrices, we are able to state our main theorem concerning solutions of the Einstein-Vlasov system with a magnetic field:

\begin{thm}\label{phystheorem}
Let $ f_0 \geq 0 $ be a smooth function with compact support in $ \bbr^3 \setminus \{ 0 \} $ and $ h \in \bbr^3 $. Suppose that $ f_0 $ is not identically zero and satisfies the constraints \eqref{mconstraint}, \eqref{tildeH}. Then, there exists a unique Bianchi I solution $ \tilde{a}_{ i j } , \tilde{k}_{ i j } \in C^0 ( ( 0 , T ] ; S_2(\bbr^3)) $  and  $ f \in C^1 ( ( 0 , T ] ; L^1 ( \bbr^3 ) ) $ to the massless (unrescaled) Einstein-Vlasov system with a magnetic field with an initial conformal gauge singularity . Furthermore, the solutions have the following asymptotics as $t \rightarrow 0^+$:
\begin{align}
 \tilde a _ { i j } & =  2t a _ { 0 i j }  + o ( t ^ {\frac32} ),\\
 \tilde{k}_{ij} & = 2 a _ { 0 i j } + o ( t^\frac12  ).
\end{align}
where $a _ { 0 i j }$ is a constant which only depends on $f_0$ and $h$. If in addition $ a_{0ij}h^ih^j  < \frac16 $, then we even have unique differentiable solutions $ \tilde{a}_{ i j } , \tilde{k}_{ i j } \in C^1 ( ( 0 , T ] ; S_2(\bbr^3)) $  and  $ f \in C^1 ( ( 0 , T ] ; L^1 ( \bbr^3 ) ) $ with the following asymptotics as $t \rightarrow 0^+$:
\begin{align}
 \tilde a _ { i j } & =  2t a _ { 0 i j }  + o ( t ^ 2 ),\\
 \tilde{k}_{ij} & = 2 a _ { 0 i j } + o ( t  ).
\end{align}
\end{thm}

The proof will be obtained using the conformally rescaled variables which will be done in the next section. Concerning solutions of the Einstein-Boltzmann system with a magnetic field our main result is as follows:

\begin{thm}\label{phystheoremboltz}
Let $ f_0 \geq 0 $ be a smooth function with compact support in $ \bbr^3 \setminus \{ 0 \} $. Suppose that $ f_0 $ is not identically zero and satisfies the constraints \eqref{mconstraint},\eqref{Hconstraint}. Then, there exists a unique Bianchi I solution $ \tilde{a}_{ i j } , \tilde{k}_{ i j } \in C^0 ( ( 0 , T ] ; S_2(\bbr^3) ) $ and $ 0 \leq f \in C^1 ( ( 0 , T ] ; L^1 ( \bbr^3 ) ) $ to the massless (unrescaled) Einstein-Boltzmann system with a magnetic field and an initial conformal gauge singularity for the scattering cross-sections in \eqref{13} such that $ f $ converges to $ f_0 $ in $ L^1 $ as $ t \to 0^+ $.  Furthermore, the solutions have the following asymptotics as $t \rightarrow 0^+$:
\begin{align}
 \tilde a _ { i j } & =  \mathcal{A} _ {  i j } t + \mathcal{B}_{ij} t  ^ { \frac { \gamma + 1 } {2 }} + o (  t^{\frac { \gamma + 1 } {2}} ),\\
 \tilde{k}_{ij} & =  \mathcal{A} _ {  i j } +  \frac { \gamma + 1 }{2} \mathcal{B}_{ij} t  ^ { \frac { \gamma - 1 } {2} } + o (t^{ \frac { \gamma - 1 } {2} } ),
\end{align}
with 
 \begin{align}
 \mathcal{A}_{ij} = 2 a_{0ij},\quad  \mathcal{B}_{ij} = \frac{1}{\gamma-1} 2^{\frac{\gamma+1}{2}} a_{1ij},
     \end{align}
     where  $a _ { 0 i j }$ and $a _ { 1 i j }$ are constants which only depend on $f_0$ and $h$. If in addition $ a_{0ij}h^ih^j < \frac16 $, there even exists a unique differentiable Bianchi I solution $ \tilde{a}_{ i j } , \tilde{k}_{ i j } \in C^1 ( ( 0 , T ] ; S_2(\bbr^3) ) $ with asymptotics as $t \rightarrow 0^+$:
\begin{align}
 \tilde a _ { i j } & =  \mathcal{A} _ {  i j } t + \mathcal{B}_{ij} t  ^ { \frac { \gamma + 1 } 2 } + \mathcal{C}_{ij} t  ^ \gamma + o ( t ^ \gamma ),\\
  \tilde{k}_{ij} & =  \mathcal{A} _ {  i j } +  \frac { \gamma + 1 } 2 \mathcal{B}_{ij} t  ^ { \frac { \gamma - 1 } 2 } + { \gamma} \mathcal{C}_{ij} t  ^ {\gamma-1} + o ( t ^ {\gamma-1} ),
\end{align}
with
 \begin{align}
\mathcal{C}_{ij}  = \frac{1}{(\gamma-1)^2} 2^{\gamma} a_{2ij},
     \end{align}
     where $a _ { 2 i j } $ is a constant which depends only on $f_0$ and $h$.  
\end{thm}

As in the Vlasov case the proof will be obtained using the conformally rescaled variables which will be done Section 5.

\medskip

\section{The Einstein-Vlasov system with a magnetic field}

Let us consider the Vlasov case where 
\begin{align}\label{Vlasov}
\frac{\partial f}{\partial \tau}=0,
\end{align}
and introduce as in \cite{LNST} the following tensor with an arbitrary number $n \geq 0 $ of indices:
\begin{align}\label{Psi}
\Psi_{i_1 i_2 \cdots i_n}= \frac { 4 \pi } { \sqrt {  a } } \int \left ( b ^ { k l } p_k p_l \right)^{ -\frac{n-1}{2} } p_{i_1} p_{i_2}\cdots p_{i_n} f  \, d^3 p,
\end{align}
which is symmetric under permutation of any of its indices and
\begin{align}
b^{ij} \Psi_{ijk_1\cdots k_m}  = \Psi_{k_1\cdots k_m},
\end{align}
for any $m\geq 2$. 

Then the main equations can be written as follows: 

\begin{align}
\label{EV1} \frac{da_{ij}}{d\tau} &= k_{ij}, \\
\frac{db^{ij}}{d\tau} & =  -b^{im}b^{jn}k_{mn},\\
\frac{dk_{ij}}{d\tau}&=-\left(\frac{2}{\tau}+\frac{b^{mn}k_{mn}}{2}\right)k_{ij}+\frac{2}{\tau}Z_{ij}-\frac{3}{\tau} \pi_{ij}^{mn} k_{mn}+k_{im}b^{mn}k_{nj}+2\Lambda \tau^2 a_{ij},\\
\nonumber \frac{d Z_{ij}}{d\tau}&=\frac{1}{\tau} \left( -Z_{ij}-k_{ij}+  \chi^{mn}_{ij} k_{mn}- \Pi_{ij}^{mn} k_{mn}\right)\\
\nonumber             & +\frac{2}{a\tau} \Big[a_{ij} h^mh^n+a_{rs}\delta^{mn}_{ij} h^rh^s-a_{ir}\delta^m_j h^rh^n-a_{is}\delta^m_j h^nh^s-a_{js}\delta^m_i h^n h^s-a_{jr}\delta^m_i h^rh^n \\
  \label{EV4}                             &+(a_{ir}a_{js}b^{mn}+a_{is}a_{jr}b^{mn}-a_{ij}a_{rs}b^{mn})h^rh^s \Big] k_{mn},
\end{align}
where
\begin{align}\label{4tensors}
\delta^{mn}_{ij}= \delta^{(m}_i\delta^{n)}_j, \quad \pi_{ij}^{mn} = \frac13 a_{ij} b^{mn},  \quad \chi^{pq}_{ij}  = b^{mp}b^{nq}\Psi_{ijmn} \quad \Pi_{ij}^{mn} = \Psi_{ij} b^{mn}.
\end{align}

$\pi_{ij}^{mn}$ is a projection since
\begin{align}
\pi_{ij}^{mn} \pi^{ij}_ {lp} = \frac13 a_{ij} b^{mn} \frac13 a_{lp} b^{ij} =  \frac13 a_{lp} b^{mn}  = \pi_{lp}^{mn}.
\end{align}
This time in contrast to the case treated in \cite{LNST} $\pi$ doesn't commute with $\chi$, since
\begin{align}
\pi_{ij}^{mn} \chi^{ij}_{lp} &= \frac13  a_{ij} b^{mn} b^{ir}b^{js}\Psi_{lprs} =  \frac13   b^{mn} b^{rs}\Psi_{lprs}=  \frac13   b^{mn} \Psi_{lp},\\
 \chi^{mn}_{ij}\pi_{lp}^{ij} & =  \frac13 b^{mr}b^{ns}\Psi_{ijrs} a_{lp} b^{ij} = \frac13 a_{lp} b^{mr}b^{ns}\Psi_{rs}.
\end{align}

In matrix form we have that
\begin{align}
&\frac{d}{d\tau}  
\left(
 \begin{matrix}
 a_{ij} \\ b^{ij}
 \end{matrix}
\right) =  \left(
 \begin{matrix}
 k_{ij} \\ -b^{im}b^{jn}k_{mn} 
 \end{matrix}
 \right), \\
&\tau \frac{d}{d\tau}  
\left(
 \begin{matrix}
 k_{ij} \\ Z_{ij}
 \end{matrix}
\right) + 
 \left(
 \begin{matrix}
 2 \delta^{mn}_{ij}+ 3 \pi_{ij}^{mn} & - 2 \delta^{mn}_{ij}\\ 
\delta^{mn}_{ij}-  \chi^{mn}_{ij}   + \Pi_{ij}^{mn}   - \mathcal{M}_{ij}^{mn}   &  \delta^{mn}_{ij}\end{matrix}
 \right) \
 \left(
 \begin{matrix}
k_{mn} \\ Z_{mn}
 \end{matrix}
\right)
 = \tau \left(
 \begin{matrix}
G_{ij} \\ 0
 \end{matrix}
\right),
\end{align}
with 
\begin{align}
  \begin{split}
  \mathcal M ^ { p q r s }
  & = \frac 2 a \Big(
  b ^ { p q } h ^ { r } h ^ { s }
  + 2 b ^ { r s } h ^ { p } h ^ { q }
  + \frac 1 2 ( a _ { m n } h ^ { m } h ^ { n } ) ( b ^ { p r } b ^ { q s } + b ^ { p s } b^ { q r } -2  b ^ { p q } b ^ { r s } )
  \\ & \quad {}
  - b ^ { q r } h ^ { p } h ^ { s }
  - b ^ { p r } h ^ { q } h ^ { s }
  - b ^ { q s } h ^ { p } h ^ { r }
  - b ^ { p s } h ^ { q } h ^ { r }
  \Big) ,
  \end{split}\\
G_{ij} &= - \frac12 b^{mn}k_{mn} k_{ij} + k_{im}b^{mn}k_{nj}+2\Lambda \tau^2 a_{ij},
\end{align}
where $G_{ij}$ should not be confused with the Einstein tensor and where we have used the notation $M_{ij}^{mn}= a_{ip}a_{jq} M^{pqmn}$.

This is in the form of Theorem 4 of \cite{LNST}  where the existence and uniqueness of solutions to the following initial value problem were considered:
\begin{equation}\label{IVP}\begin{aligned}
  x ' ( \tau ) & = F ( \tau , x ( \tau ) , y ( \tau) ) ,
  \qquad & x ( 0 ) = x _ 0 ,
  \\
  \tau y ' ( \tau ) + N ( \tau , x ( \tau ) ) y ( \tau) & = \tau G ( \tau , x ( \tau ) , y ( \tau ) ) + H ( \tau , x ( \tau ) ) ,
  \qquad
&  y ( 0 ) = y _ 0 .
\end{aligned}\end{equation}
 with
\begin{align}
x = 
\left(
\begin{array}{c}
a_{ij} \\
b^{ij}
\end{array}
\right), \quad 
y = 
\left(
\begin{array}{c}
k_{ij} \\
Z_{ij}
\end{array}
\right), \quad  F= \left(
 \begin{matrix}
 k_{ij} \\ -b^{im}b^{jn}k_{mn} 
 \end{matrix}
 \right), \quad G =
\left(
\begin{array}{c}
G_{ij} \\
0
\end{array}
\right), \quad H =\left(
\begin{array}{c}
0 \\
0
\end{array}
\right),
\end{align}
and 
\begin{align}\label{N}
N=  \left(
 \begin{matrix}
 2 \delta^{mn}_{ij}+ 3 \pi_{ij}^{mn} & - 2 \delta^{mn}_{ij} \\ 
\delta^{mn}_{ij}  \chi^{mn}_{ij}   + \Pi_{ij}^{mn}   - \mathcal{M}_{ij}^{mn}   & \delta^{mn}_{ij}\end{matrix}
 \right) .
 \end{align}
We now need to check whether the conditions of Theorem 4 of \cite{LNST} are satisfied. There is the compatibility condition 
\begin{align}\label{comp}
N(0,x_0) y_0 = H (0, x_0),
\end{align}
and the differentiability condition that $F$, $G$, $H$, $N$, $\frac{\partial F}{\partial x}$, $\frac{\partial F}{\partial y}$, $\frac{\partial G}{\partial  x}$, $\frac{\partial G}{\partial y}$, $\frac{\partial H}{\partial \tau}$, $\frac{\partial H}{\partial x}$, $\frac{\partial^2 H}{\partial \tau \partial x}$, $\frac{\partial^2 H}{\partial x^2}$, $\frac{\partial N}{\partial \tau}$, $\frac{\partial N}{\partial x}$, $\frac{\partial^2 N}{\partial \tau \partial x}$, $\frac{\partial^2 N}{\partial x^2}$ should be continuous. Moreover we need that all the eigenvalues of  $N(0,x_0)$ have positive real part. If these conditions are satisfied there is a unique solution to the initial value problem \eqref{IVP} for all initial data satisfying the compatibility condition \eqref{comp} and this solution depends continuously on $F$, $G$, $H$, $N$, $x_0$ and $y_0$. The solution is continuous for $\tau \geq 0$ and is continuously differentiable and satisfies the differential equation equation for $\tau > 0$. If all the eigenvalues of $N$ have real parts which are greater than $1$ then the solution is continuously differentiable and satisfies the equation for $ \tau \geq 0$. We will obtain all these conditions. The last condition we will obtain by restricting the size of the norm of the magnetic field.

\medskip

\subsection{The differentiability condition}

In the Vlasov case there is no $H$ and that the coefficient functions $F$ and $G$ have the differentiability properties needed follows from \cite{LNST} where the case without a magnetic field was considered. The same happens with $N$ except where the term coming from the magnetic field appears. We just have to check the differentiability properties of $\mathcal{M}_{ij}^{mn}$ which can be seen to be satisfied since $\mathcal{M}_{ij}^{mn}$ is a rational function of $a_{ij}$.

\medskip

\subsection{Fuchsian condition at $ \tau = 0 $}

We will consider in this and the next subsection that all functions are evaluated at $\tau=0$  e.g.
we write ${ a _ { i j } }$ for ${ a _ { i j } ( 0 ) }$, etc.

For the variable $Z_{ij}$ defined in \eqref{19} to be well-defined at $\tau=0$ we need
\begin{align}\label{fuchs1}
2\Psi_{ij} = a_{ij} - \frac{2}{a}(a_{ij}a_{mn}-2a_{im}a_{jn})h^mh^n,
\end{align}
which is a relation between the initial data $(f, a_{ij},h^i)$. 

We will see that this condition will be satisfied for certain critical points of a functional which will be defined in the following. We begin by collecting a number of facts about the space of
positive definite symmetric bilinear forms
which we will need in our discussion of data satisfying the Fuchsian condition.

We'll take our symmetric bilinear forms to have upper indices since
that's what we need later but everything would work similarly, with
minor changes, for lower indices.

It's convenient to use ${ b ^ { i j } }$ as coordinates on the space
of positive definite symmetric forms, but ${ b ^ { i j } = b ^ { j i } }$
so these aren't really
independent coordinates. We can nonetheless treat them as if they
were as long as we are careful to fully symmetrise expressions in
which they appear and as long as we adopt a sensible convention
on the meaning of partial derivatives. In particular
${ \frac{\partial }{\partial b ^ { i j }} }$, as a vector field on the space of all
bilinear forms, is not tangential on the subspace of symmetric
bilinear forms and so cannot be applied to functions defined on
the space of symmetric bilinear forms. We will however always
consider it to be a shorthand for the vector field
\[
  \frac 1 2 \left ( \frac { \partial } {  \partial b ^ { i j } }
    + \frac {  \partial } {  \partial b ^ { j i } } \right ) ,
\]
which \emph{is} tangential on the space of symmetric bilinear forms
and so can sensibly be applied to functions on this space. With
this convention we have
\begin{equation}
  \frac {  \partial b ^ { i j } } {  \partial b ^ { k l } } = \delta ^ { i j } _ { k l },
\end{equation}
rather than having ${ \delta ^ { i } _ { k } \delta ^ { j } _ { l } }$ on
the right hand side.

We will adopt the same conventions as elsewhere in the paper as
to the meanings of ${ a _ { i j } }$, ${ a }$, and ${ b }$, namely
that ${ a _ { i j } }$ is the inverse to ${ b ^ { i j } }$ and
${ a }$ and ${ b }$ are the determinants of ${ a _ { i j } }$ and
${ b ^ { i j } }$, respectively.

The tangent space to the space of positive definite symmetric
bilinear forms is the space of all symmetric bilinear forms.
There is a natural positive definite symmetric bilinear form
on the tangent space at the point ${ b ^ { i j } }$, namely
\begin{equation}
  g _ { k l , m n } \left ( b ^ { i j } \right )
  = \frac 1 2 \left ( a _ { k m } a _ { l n } + a _ { k n } a _ { l m } \right ) .
\end{equation}
A choice of a positive definite symmetric form on the tangent space
at each point is of course a Riemannian metric, hence the notation above.
When we say the choice above is natural we mean that the metric
is independent of the choice of basis with respect to which we
take components. It is, up to scaling, the only choice of metric
with this property. It makes the space of symmetric positive
definite bilinear forms into a Riemannian symmetric space.

Now that we have equipped the space of positive definite symmetric
bilinear forms with a Riemannian metric we can do differential
geometry on it. Consider, for example, the connection
\begin{equation}
  \Gamma ^ { k l } _ { m n , p q } \left ( b ^ { i j } \right )
  = - \frac 1 4 \left (
    a _ { m p } \delta ^ { k l } _ { n q }
    + a _ { m q } \delta ^ { k l } _ { n p }
    + a _ { n p } \delta ^ { k l } _ { m q }
    + a _ { n q } \delta ^ { k l } _ { m p }
  \right ) .
\end{equation}
Easy calculations show that
\begin{equation}
  \nabla _ { i j } a _ { k l } = 0 , \qquad \nabla _ { i j } b ^ { k l } = 0 ,
\end{equation}
where ${ \nabla_ { i j } }$ is the covariant derivative in the ${ b ^ { i j } }$
direction,
from which it follows that
\begin{equation}
  \nabla _ { i j } g _ { k l , m n } = 0 .
\end{equation}
Since the covariant derivative of the metric is zero and the connection
is clearly torsion free we conclude that it is the Levi-Civita connection.
This can of course also be verified directly from the usual formula for
the Levi-Civita connection in terms of derivatives of the metric
coefficients.
Another useful relation is
\begin{equation}
  \nabla _ { i j } b = b a _ { i j } ,
\end{equation}
since the covariant derivative of a scalar is just the ordinary derivative.
It is important to remember that tensors which appear to be constant,
i.e. have the same components at all points, need not have covariant
derivative zero. For example, if the components of ${ \eta ^ { i j } }$
are constant then
\begin{equation}
  \begin{split}
    \nabla _ { k l } \eta ^ { i j }
    & = \Gamma ^ { i j } _ { k l , m n } \eta ^ { m n }
    \\ & = - \frac 1 4 \left (
      a _ { k m } \delta ^ { i j } _ { l n } \eta ^ { m n }
      + a _ { k n } \delta ^ { i j } _ { l m } \eta ^ { m n }
      + a _ { l m } \delta ^ { i j } _ { k n } \eta ^ { m n }
      + a _ { l n } \delta ^ { i j } _ { k m } \eta ^ { m n }
    \right )
    \\ & = - \frac 1 8 \Big (
      a _ { k m } \delta ^ { i } _ { l } \eta ^ { j m }
      + a _ { k m } \delta ^ { j } _ { l } \eta ^ { i m }
      + a _ { k n } \delta ^ { i } _ { l } \eta ^ { j n }
      + a _ { k n } \delta ^ { j } _ { l } \eta ^ { i n }
    \\ & \qquad {}
      + a _ { l m } \delta ^ { i } _ { k } \eta ^ { j m }
      + a _ { l m } \delta ^ { j } _ { k } \eta ^ { i m }
      + a _ { l n } \delta ^ { i } _ { k } \eta ^ { j n }
      + a _ { l n } \delta ^ { j } _ { k } \eta ^ { i n }
    \Big ) .
  \end{split}
\end{equation}

The differential equation for geodesics is
\begin{equation}
  \frac { d ^ 2 b ^ { i j } } { d s ^ 2 }
  + \Gamma ^ { i j } _ { k l , m n } \frac { d b ^ { k l } } { d s }
    \frac { d b ^ { m n } } { d s } = 0 .
\end{equation}
It is easily seen that the matrix valued function
\[
  L ^ T \exp ( s D ) L ,
\]
where ${ D }$ is a diagonal matrix and ${ L }$ is an invertible
matrix, is a solution. These are in fact the only solutions,
since we can choose ${ L }$ and ${ D }$ to match any initial
conditions and we know the differential equation has a unique
solution for any choice of initial conditions. Clearly all
geodesics are infinitely extensible so the space of symmetric
positive definite bilinear forms is geodesically complete.
Given a pair of distinct forms we can find a change of basis
matrix ${ L }$ which simultaneously diagonalises them
and such that one of them has the identity matrix as coefficient
matrix. Taking the logarithm of the other as ${ D }$ we find
that the geodesic given above passes through both, at the
parameter values ${ s = 0 }$ and ${ s = 1 }$. This is in
fact the only geodesic through those points, so the
space of symmetric positive definite bilinear forms is
geodesically convex.

Define
\begin{equation}
  F ( b ^ { i j } , h ^ { i } , f )
  = 4 b ^ { - 1 / 2 } \Psi + 2 b ^ { - 1 / 2 } + 4 b ^ { 1 / 2 } a _ { k l } h ^ { k } h ^ { l } .
\end{equation}
Here ${ b ^ { i j } }$ ranges over symmetric ${ 3 \times 3 }$ matrices,
${ h ^ { i } }$ ranges over ${ \mathbb{ R} ^ 3 }$ and ${ f }$ ranges
over non-negative functions on ${ \mathbb{ R} ^ 3 }$ for which the integral
\[
  \int \left ( \delta ^ { i j } p _ i p _ j \right ) ^ { 1 / 2 }
  f ( p ) \, d ^ 3 p,
\]
is finite and positive. The particular symmetric bilinear form ${ \delta ^ { i j } }$ appears to play a distinguished role here but it is easy to see that it could be replaced here by any other
choice of positive definite symmetric bilinear form. 
In order to compute the covariant derivative of $F$ we observe that 
\begin{equation}
  4 b ^ { - 1 / 2 } \Psi = 16 \pi \int \left ( b ^ { m n } p _ m p _ n \right ) ^ { 1 / 2 } f \, d ^ 3 p ,
\end{equation}
so that
\begin{align}
  \nonumber \nabla _ { k l } \left ( 4 b ^ { - 1 / 2 } \Psi \right )& = \partial_ { k l } \left ( 4 b ^ { - 1 / 2 } \Psi \right )
    = 16 \pi \int \partial_ { k l } \left ( b ^ { m n } p _ m p _ n \right ) ^ { 1 / 2 } f \, d ^ 3 p \\
   & = 8 \pi \int \left ( b ^ { m n } p _ m p _ n \right ) ^ { - 1 / 2 } p _ k p _ l f \, d ^ 3 p = 2 b ^ { - 1 / 2 } \Psi _ { k l } , \label{cov1}
\end{align}
and
\begin{align}
 \nonumber   \nabla _ { k l } \left ( 4 b ^ { 1 / 2 } a _ { m n } h ^ m h ^ n \right ) &= \nabla _ { k l } \left ( 4 b ^ { 1 / 2 } \right ) a _ { m n } h ^ m h ^ n
      + b ^ { 1 / 2 } a _ { m n } \nabla _ { k l } \left ( 4 h ^ m h ^ n \right )
    \\  &= 2 b ^ { 1 / 2 } \left (a _ { k l } a _ { m n } - a _ { k m } a _ { l n } - a _ { k n } a _ { l m }\right ) h ^ m h ^ n. \label{cov2}
\end{align}
We thus obtain
\begin{equation}
  \nabla _ { i j } F = \frac { \partial F } { \partial b ^ { i j } }
  = b ^ { - 1 / 2 } \left [ 2 \Psi _ { i j } - a _ { i j }
    + \frac 2 a \left ( a _ { i j } a _ { m n } - a _ { i m } a _ { j n }
    - a _ { i n } a _ { j m } \right ) h ^ { m } h ^ { n } \right ] .
\end{equation}
It follows that the ${ b ^ { i j } }$ for which ${ F }$ has a
critical point, for fixed choices of ${ h ^ { i } }$ and ${ f }$,
are precisely the ones which satisfy the Fuchsian condition \eqref{fuchs1}.

Suppose that ${ b ^ { i j } }$ and ${ C }$ are such that
\begin{equation}
  F ( b ^ { i j } , f , h ^ { i } ) \le C.
\end{equation}
Each term in the definition of ${ F }$ is non-negative so it follows that
\begin{equation}
  4 b ^ { - 1 / 2 } \Psi \le C,
\end{equation}
and
\begin{equation}
  2 b ^ { - 1 / 2 } \le C .
\end{equation}
From the latter it follows that
\begin{equation}
  b \ge \frac 4 { C ^ 2 } .
\end{equation}
Let ${ b _ { \mathrm{max} } }$ be the largest eigenvalue of ${ b ^ { i j } }$,
relative to ${ \delta ^ { i j } }$. Again, we could, if we wanted replace ${ \delta ^ { i j } }$ by any other positive definite symmetric bilinear form, provided we replace the dot products and norms below by the corresponding products and norms defined in terms of that form. There is nothing to be gained by doing so however, other than the observation that the space of
symmetric positive definite bilinear forms has, as far as this section is concerned, no distinguished point.

Now choose ${ \xi ^ { i } }$ to be an eigenvector with that eigenvalue,
normalised to have Euclidean norm~1.
Then
\begin{equation}
  b _ { \mathrm{max} } | \xi \cdot p | ^ 2 \le b ^ { k l } p _ { k } p _ { l },
\end{equation}
for all ${ p }$ and hence
\begin{equation}
  16 \pi  b _ { \mathrm{max} } ^ { 1 / 2 }
  \int | \xi \cdot p | f ( p ) d ^ 3 p
  \le 16 \pi \int \left ( b ^ { k l } p _ { k } p _ { l } \right ) ^ { 1 / 2 }
    f ( p ) d ^ 3 p = 4 b ^ { - 1 / 2 } \Psi \le C .
\end{equation}
Since ${ \xi }$ has norm~1 we obtain
\begin{equation}
  16 \pi b _ { \mathrm{max} } ^ { 1 / 2 } \min _ { \| \xi \| = 1 }
  \int | \xi \cdot p | f ( p ) d ^ 3 p
  \le C .
\end{equation}
Under our hypotheses on ${ f }$ this minimum exists and is positive so
\begin{equation}
  b _ { \mathrm{max} } ^ { 1 / 2 }
  \le \frac { C } { 16 \pi  \min _ { \| \xi \| = 1 } \int | \xi \cdot p | f ( p ) d ^ 3 p } .
\end{equation}
Let ${ R _ C }$ be the set of symmetric ${ 3 \times 3 }$ matrices ${ b ^ { i j } }$ satisfying the inequalities
\begin{equation}
  b \ge \frac 4 { C ^ 2 } , \quad
  b _ { \mathrm{max} } \le \frac { C ^ 2 } { 256 \pi ^ 2 \min _ { \| \xi \| = 1 } \left ( \int | \xi \cdot p | f ( p ) d ^ 3 p \right ) ^ 2 } .
\end{equation}
Of course ${ R _ C }$ depends not just on ${ C }$ but also on ${ f }$, although it doesn't depend on ${ h ^ i }$.
What we've just shown is that any ${ b ^ { i j } }$ for which ${ F }$ is at most ${ C }$ lie in ${ R _ C }$ or, equivalently, that outside of ${ R _ C }$ the value of ${ F }$ is always greater than~${ C }$.

The set ${ R _ C }$ defined above is compact.
It could be empty but at least for
${ C = F ( \delta ^ { i j } , h ^ { i } , f ) }$
it is not, since it contains~${ \delta ^ { i j } }$.
Since ${ F }$ is continuous it follows that for any
${ h ^ { i } }$ and ${ f }$ the function ${ F }$ has a
minimum on ${ R _ C }$ for this choice of ${ C }$,
which can be at most ${ C }$ since we have a point inside
${ R _ C }$ where it takes the value ${ C }$.
But then it has a global minimum since we've already seen
that its values outside ${ R _ C }$ are greater than ${ C }$.
This global minimum must be a critical point, and hence
a solution of the Fuchsian conditions.

Compared to existence, the uniqueness of solutions to the Fuchsian conditions is relatively difficult. Here we use the Riemannian structure on the space of positive definite
symmetric bilinear forms discussed above.

The covariant Hessian of ${ 4 b ^ { - 1 / 2 } \Psi }$ is computed using \eqref{cov1} as follows:
\begin{equation}
  \begin{split}
    \nabla _ { i j } \nabla _ { k l } \left ( 4 b ^ { - 1 / 2 } \Psi \right )
    & = \partial_ { i j } \nabla _ { k l } \left ( 4 b ^ { - 1 / 2 } \Psi \right )
        - \Gamma ^ { p q } _ { i j , k l } \nabla _ { p q } \left ( 4 b ^ { - 1 / 2 } \Psi \right )
    \\ & = - 4 \pi \int \left ( b ^ { m n } p _ m p _ n \right ) ^ { - 3 / 2 } p _ i p _ j p _ k p _ l f \, d ^ 3 p
    \\ & \quad {}
      + \frac 1 2 b ^ { - 1 / 2 } \left (
        a _ { i k } \Psi _ { j l }
        + a _ { i l } \Psi _ { j k }
        + a _ { j k } \Psi _ { i l }
        + a _ { j l } \Psi _ { i k }
     \right )
    \\ & = \frac 1 2 b ^ { - 1 / 2 } \left (
        - 2 \Psi _ { i j k l }
        + a _ { i k } \Psi _ { j l }
        + a _ { i l } \Psi _ { j k }
        + a _ { j k } \Psi _ { i l }
        + a _ { j l } \Psi _ { i k }
     \right ) .
  \end{split}
\end{equation}
This is positive definite. In other words, for any non-zero symmetric
bilinear form ${ c ^ { i j } }$ we have
\begin{equation}
    \nabla _ { i j } \nabla _ { k l } \left ( 4 b ^ { - 1 / 2 } \Psi \right )
    c ^ { i j } c ^ { k l } > 0 .
\end{equation}
To see this we write the left hand side as the integral of
the function
\[
  \left \{ \left [
    b ^ { m n } p _ m p _ n \left (
      a _ { i k } p _ j p _ l + a _ { i l } p _ j p _ k
      + a _ { j k } p _ i p _ l + a _ { j l } p _ i p _ k
    \right ) \right ]
    - 2 p _ i p _ j p _ k p _ l \right \}
    c ^ { i j } c ^ { kl },
\]
with respect to the measure
\[
  2 \pi ( b ^ { m n } p _ m p _ n ) ^ { - 3 / 2 }
    f \, d ^ 3 p .
\]
This function is invariant under linear changes of variable and so we may make such a change to make the matrix representing ${ b }$
the identity matrix and the identity representing ${ c }$ a diagonal matrix, whose three diagonal entries we'll
call ${ \xi_ 1 }$, ${ \xi _ 2 }$, and ${ \xi _ 3 }$. With these choices
the integrand becomes
\[
  4 \left ( p _ 1 ^ 2 + p _ 2 ^ 2 + p _ 3 ^ 2 \right )
    \left ( \xi _ 1 ^ 2 p _ 1 ^ 2 + \xi _ 2 ^ 2 p _ 2 ^ 2 + \xi_ 3 ^ 2 p _ 3 ^ 2 \right )
  - 2 \left ( \xi _ 1 p _ 1 ^ 2 + \xi _ 2 p _ 2 ^ 2 + \xi _ 3 p _ 3 ^ 2 \right ) ^ 2 .
\]
As a consequence of the Cauchy-Schwarz inequality this is bounded from below
by
\[
  2 \left ( p _ 1 ^ 2 + p _ 2 ^ 2 + p _ 3 ^ 2 \right )
    \left ( \xi _ 1 ^ 2 p _ 1 ^ 2 + \xi _ 2 ^ 2 p _ 2 ^ 2 + \xi _ 3 ^ 2 p _ 3 ^ 2 \right ) ,
\]
which is clearly non-negative and which, because of our assumption that
${ c ^ { i j } }$ is non-zero, is positive off of a point, line, or
plane, depending on the number of zero eigenvalues of ${ c }$ relative
to ${ a }$. In particular there
must be a set of positive measure, with respect to the measure
\[
  2 \pi ( b ^ { m n } p _ m p _ n ) ^ { - 3 / 2 }
    f \, d ^ 3 p ,
\]
where the integrand is positive. It follows that
\begin{equation}
    \nabla _ { i j } \nabla _ { k l } \left ( 4 b ^ { - 1 / 2 } \Psi \right )
    c ^ { i j } c ^ { k l } > 0 ,
\end{equation}
as claimed.
In other words, ${ 4 b ^ { - 1 / 2 } \Psi }$ is strictly geodesically
convex.

We can easily check that the middle term in ${ F }$ is geodesically convex.
It suffices to note that
\begin{equation}
  \nabla _ { i j } \nabla _ { k l } \left ( 2 b ^ { - 1 / 2 } \right )
  = \nabla_ { i j } \left ( - b ^ { - 1 / 2 } a _ { k l } \right )
  = \frac 1 2 b ^ { - 1 / 2 } a _ { i j } a _ { k l } .
\end{equation}

Finally, we have to consider the last term in ${ F }$.  Using \eqref{cov2} we obtain
\begin{equation}
  \begin{split}
    \nabla _ { i j } \nabla _ { k l } \left ( 4 b ^ { 1 / 2 } a _ { m n } h ^ m h ^ n \right )
    & = \nabla _ { i j } \left [
      2 b ^ { 1 / 2 } \left (
        a _ { k l } a _ { m n }
        - a _ { k m } a _ { l n }
        - a _ { k n } a _ { l m }
      \right ) h ^ m h ^ n
    \right ]
    \\ & = \nabla_ { i j } \left ( 2 b ^ { 1 / 2 } \right )
      \left (
        a _ { k l } a _ { m n }
        - a _ { k m } a _ { l n }
        - a _ { k n } a _ { l m }
      \right ) h ^ m h ^ n
    \\ & \quad {}
      + 
      b ^ { 1 / 2 } \left (
        a _ { k l } a _ { m n }
        - a _ { k m } a _ { l n }
        - a _ { k n } a _ { l m }
      \right ) \nabla _ { i j } \left ( 2 h ^ m h ^ n \right )
    \\ & = b ^ { 1 / 2 } \Big (
        a _ { i j } a _ { k l } a _ { m n }
        - a _ { i j } a _ { k m } a _ { l n }
        - a _ { i j } a _ { k n } a _ { l m }
    \\ & \quad {}
        - a _ { i m } a _ { j n } a _ { k l }
        - a _ { i n } a _ { j m } a _ { k l }
        + \frac 1 2 a _ { i k } a _ { j m } a _ { l n }
    \\ & \quad {}
        + \frac 1 2 a _ { i k } a _ { j n } a _ { l m }
        + \frac 1 2 a _ { i l } a _ { j m } a _ { k n }
        + \frac 1 2 a _ { i l } a _ { j n } a _ { k m }
    \\ & \quad {}
        + \frac 1 2 a _ { i m } a _ { j k } a _ { l n }
        + \frac 1 2 a _ { i m } a _ { j l } a _ { k n }
        + \frac 1 2 a _ { i n } a _ { j k } a _ { l m }
    \\ & \quad {}
        + \frac 1 2 a _ { i n } a _ { j l } a _ { k m }
      \Big ) h ^ m h ^ n.
  \end{split}
\end{equation}
We can see that this is positive definite as follows.
We apply it to an arbitrary symmetric bilinear
form ${ c ^ { i j } }$, obtaining a positive factor of
${ b ^ { 1 / 2 } }$ times
\[
  \begin{split}
    & \Big (
        a _ { i j } a _ { k l } a _ { m n }
        - a _ { i j } a _ { k m } a _ { l n }
        - a _ { i j } a _ { k n } a _ { l m }
        - a _ { i m } a _ { j n } a _ { k l }
        - a _ { i n } a _ { j m } a _ { k l }
    \\ & \quad {}
        + \frac 1 2 a _ { i k } a _ { j m } a _ { l n }
        + \frac 1 2 a _ { i k } a _ { j n } a _ { l m }
        + \frac 1 2 a _ { i l } a _ { j m } a _ { k n }
        + \frac 1 2 a _ { i l } a _ { j n } a _ { k m }
    \\ & \quad {}
        + \frac 1 2 a _ { i m } a _ { j k } a _ { l n }
        + \frac 1 2 a _ { i m } a _ { j l } a _ { k n }
        + \frac 1 2 a _ { i n } a _ { j k } a _ { l m }
        + \frac 1 2 a _ { i n } a _ { j l } a _ { k m }
      \Big ) c ^ { i j } c ^ { k l } h ^ m h ^ n .
  \end{split}
\]
As we did for the first term, we evaluate this in a basis in which
${ b }$ is represented by the identity matrix and ${ c }$ is represented
by a diagonal matrix with diagonal entries ${ \xi _ 1 }$, ${ \xi _ 2 }$
and ${ \xi_ 3 }$, obtaining
\[
  ( - \xi _ 1 + \xi_ 2 + \xi _ 3 ) ^ 2 h _ 1 ^ 2
  + ( \xi_ 1 - \xi _ 2 + \xi _ 3 ) ^ 2 h _ 2 ^ 2
  + ( \xi _ 1 + \xi _ 2 - \xi _ 3 ) ^ 2 h _ 3 ^ 2 .
\]
This is clearly non-negative so
\[
  \nabla_ { i j } \nabla _ { k l } \left ( 4 b ^ { 1 / 2 } a _ { m n } h ^ m h ^ n \right ),
\]
is positive semidefinite and ${ 4 b ^ { 1 / 2 } a _ { m n } h ^ m h ^ n }$
is geodesically convex.

We've now seen that ${ F }$ is a sum of three terms, the first of which
is strictly geodesically convex and the other two of which are
geodesically convex. It follows that ${ F }$ is strictly geodesically convex.
Suppose it had two distinct critical points. We've already seen that there must
be a geodesic connecting them. The restriction of ${ F }$ to this
geodesic would be a strictly convex function with at least two critical
points. This is impossible, so ${ F }$ has at most one critical point on
the set of positive definite symmetric bilinear forms. We've already
seen that it has at least one, which is a minimum, so there is a unique
critical point. In other words, for any allowed
choice of ${ f }$ and ${ h ^ i }$
there is a unique solution of the Fuchsian conditions.

In the special case ${ h = 0 }$ this was stated in \cite{AT2}. The proof of uniqueness given there relies on the claim that a continuously differentiable function on a contractible set, all of
whose critical points are minima, has at most one critical point. Unfortunately 
\begin{align*} 17 y ^ 4 + 8 x y ^ 2 + x ^ 2 - 17 y ^ 2 - 4 x 
\end{align*}
 has exactly three critical points, at ${ ( 0 , \sqrt { 1 / 2 } ) }$, ${ ( 0 , - \sqrt { 1 / 2 } ) }$ and ${ ( 2 , 0 ) }$. The first two of these are the only ones in the unit disc, which is a contractible set, and both are minima. So their argument is incorrect, but the argument given here applies to
the case ${ h = 0 }$ and it can serve as a replacement for the one given there.

Now, given $(f,h^i,a_{ij})$ we obtain from \eqref{comp} for $\tau =0$ that
\begin{align}\label{compVlasov}
\left(
 \begin{matrix}
 2 \delta^{mn}_{ij}+ 3 \pi_{ij}^{mn} & - 2 \delta^{mn}_{ij}\\ 
\delta^{mn}_{ij}-  \chi^{mn}_{ij}   + \Pi_{ij}^{mn}   - \mathcal{M}_{ij}^{mn}   &  \delta^{mn}_{ij}\end{matrix}
 \right) \
 \left(
 \begin{matrix}
k_{mn} \\ Z_{mn}
 \end{matrix}
\right)
 =  \left(
 \begin{matrix}
0 \\ 0
 \end{matrix}
\right).
\end{align}
We will show in the next section that all the eigenvalues of $N$ in \eqref{N} have positive real part and $N$ is thus invertible. In that case we have from \eqref{compVlasov} that there are unique $k_{ij}$, $Z_{ij}$ which in fact have to vanish initially. Having in mind that we will show that eigenvalue condition in the next section,  let us summarise the result of this section as follows:

\begin{prop}\label{propVlasov}
 Let $f_0 \geq 0$ be a smooth function with compact support in $\mathbb{R}^3\setminus\{ 0 \}$. Suppose that
$f_0$ is not identically zero. Then, there exist unique $ 3 \times 3$ symmetric matrices $a_0$, $b_0$, $k_0$ and $Z_0$ and a vector $h$ satisfying the Fuchsian conditions \eqref{fuchs1} and  \eqref{compVlasov}.
 \end{prop}
 
 From \eqref{fuchs1} we may deduce also for $\tau =0$ that

\begin{align}\label{fuchs11}
2 \chi ^{mn}_{ij} b^{ij} = & 2\Psi_{ij} b^{im}b^{jn} = b^{mn}\left( 1 - \frac{2}{a} a_{rs}h^r h^s  \right)  + \frac{4}{a} h^m h^n,\\
\label{fuchs12}\mathcal{M}_{ij}^{mn}b^{ij} & =- \frac{2 }{a } h^mh^n.
\end{align}

Let us consider two vectors $e^i$, $f^i$ which are orthogonal to each other and to the magnetic field, such that:
\begin{align}\label{basisvectors}
a_{ij} e^i f^j = a_{ij} e^i h^j = a_{ij} f^i h^j = 0,\quad a_{ij}e^i e^j = a_{ij}f^i f^j.
\end{align}

Contracting \eqref{fuchs1} with  $e^i e^j$
\begin{align}
2\Psi_{ij} e^{i}e^{j} = a_{ij}e^i e^j  (1- \frac{2}{a} a_{mn}h^m h^n),
\end{align}
which since $\Psi_{ij} e^{i}e^{j}$ and $a_{ij} e^{i}e^{j}$ are both positive definite, implies 
\begin{align} \label{h2bound}
a_{mn}h^m h^n  < \frac{a}{2}.
\end{align}

Thus, although $h^i$ was freely specifiable, once we have found $a_{ij}$ satisfying \eqref{fuchs1}, the norm of $h^i$ is bounded above.

\medskip

\subsection{The eigenvalues of $N$ at $\tau=0$}

We follow \cite{AT2} and consider the eigenvalues of $\chi$ which we name $\mu$, i.e.  $\chi^{pq}_{ij} k^{ij}=\mu k^{pq}$. Working in the basis where $a_{ij}= \delta_{ij}$ and $k_{ij}=\diag (k_1,k_2,k_3)$ we obtain
\begin{align}
\nonumber \vert  \mu  k_i \vert =   \left \vert 4 \pi  \int \left ( \delta ^ { k l } p_k p_l \right)^{ -\frac{3}{2} } p_{i}^2 (p_1^2 k_1 + p_2^2 k_2+p_3^2 k_3) f  \, d^3 p \right \vert\\
\leq 4 \pi   \int \left ( \delta ^ { k l } p_k p_l \right)^{ -\frac{3}{2} } p_{i}^2 (p_1^2 \vert k_1 \vert+ p_2^2  \vert  k_2  \vert +p_3^2  \vert k_3  \vert ) f d^3 p.
\end{align}
This implies
\begin{align}
\nonumber \sum_{i=1}^3 \vert  \mu  k_i \vert &\leq 4 \pi   \int \left ( \delta ^ { k l } p_k p_l \right)^{ -\frac{1}{2} }  (p_1^2 \vert k_1 \vert+ p_2^2  \vert  k_2  \vert +p_3^2  \vert k_3  \vert ) f  = \Psi_{11} \vert k_1 \vert + \Psi_{22} \vert k_2 \vert  + \Psi_{33} \vert k_3 \vert\\
\nonumber& =  \left[\frac12 - \delta_{mn}h^m h^n + 2(h^1)^2  \right] \vert k_1 \vert  + \left[\frac12 -  \delta_{mn}h^m h^n + 2(h^2)^2  \right] \vert k_2 \vert +\left [\frac12 -  \delta_{mn}h^m h^n + 2(h^3)^2   \right]  \vert k_3 \vert \\
\nonumber & =  \left(\frac12 -  \delta_{mn}h^m h^n  \right) \sum_{i=1}^3   \vert k_i \vert + 2(h^1)^2 \vert k_1 \vert +  2(h^2)^2 \vert k_2 \vert + 2(h^3)^2  \vert k_3 \vert \\ 
& \leq \left(\frac12 +  \delta_{mn}h^m h^n \right) \sum_{i=1}^3 \vert k_i \vert,
\end{align}
which implies 
\begin{align}\label{mu}
\vert \mu \vert < \frac12 + \delta_{mn}h^m h^n.
\end{align}

We also have
\begin{align}
2 \Pi^{mn}_{ij} = 3 \pi^{mn}_{ij} - \frac{2}{a} b^{mn} (a_{ij}a_{rs}h^r h^s -2a_{ir}h^ra_{js}h^s) = \left( 3 - \frac{6}{a} a_{rs}h^r h^s \right)  \pi^{mn}_{ij} + \frac{12}{a}  \pi^{mn}_{ir} h^r a_{js}h^s.
\end{align}

Denote by $\hat{k}_{ij}$ and $\hat{Z}_{ij}$ components of an eigenvector of $N$. The eigenvalue equations for $N$ using $\lambda$ for the eigenvalue are

\begin{align}
2 \hat{k}_{ij}+ 3 \pi_{ij}^{mn} \hat{k}_{mn}  - 2 \hat{Z}_{ij} = \lambda \hat{k}_{ij}, \\
\hat{k}_{ij}-  \chi^{mn}_{ij}\hat{k}_{mn}   + \Pi_{ij}^{mn} \hat{k}_{mn}   - \mathcal{M}_{ij}^{mn}  \hat{k}_{mn} + \hat{Z}_{ij} = \lambda \hat{Z}_{ij}.
\end{align}
From the first equation we have
\begin{align}
\hat{Z}_{ij} = \hat{k}_{ij}+ \frac32 \pi_{ij}^{mn} \hat{k}_{mn} - \frac12 \lambda \hat{k}_{ij},
\end{align}
so that the second equation becomes
\begin{align}
\hat{k}_{ij}-  \chi^{mn}_{ij}\hat{k}_{mn}   + \Pi_{ij}^{mn} \hat{k}_{mn}   - \mathcal{M}_{ij}^{mn}  \hat{k}_{mn}  = (\lambda -1)  \left( \hat{k}_{ij}+ \frac32 \pi_{ij}^{mn} \hat{k}_{mn} - \frac12 \lambda \hat{k}_{ij} \right),
\end{align}
which using the definitions of $\chi^{mn}_{ij}$, $\Pi_{ij}^{mn}$ and $\pi_{ij}^{mn}$ \eqref{4tensors} turns into
\begin{align*}
 \hat{k}_{ij}-  b^{mr}b^{ns}\Psi_{ijrs}  \hat{k}_{mn}   + \Psi_{ij} b^{mn}\hat{k}_{mn}   - \mathcal{M}_{ij}^{mn}  \hat{k}_{mn}  
=  (\lambda -1) \left( \hat{k}_{ij}+ \frac12 a_{ij} b^{mn} \hat{k}_{mn} - \frac12 \lambda \hat{k}_{ij}\right),
\end{align*}
and using the notation $\hat{k}=b^{mn}\hat{k}_{mn}$:
\begin{align} \label{sec}
\hat{k}_{ij}-  b^{mr}b^{ns}\Psi_{ijrs}  \hat{k}_{mn}   + \Psi_{ij} \hat{k}   - \mathcal{M}_{ij}^{mn}  \hat{k}_{mn}
= (\lambda -1) \left( \hat{k}_{ij}+ \frac12 a_{ij}  \hat{k} - \frac12 \lambda \hat{k}_{ij}\right).
\end{align}

If one contracts the last equation with $b^{ij}$ and uses \eqref{fuchs1}, \eqref{fuchs11}-\eqref{fuchs12} one obtains
\begin{align}
\nonumber \hat{k} -  \frac12 \left( 1 - \frac{2}{a} a_{rs}h^r h^s \right)  \hat{k} - \frac{2}{a} h^m h^n  \hat{k}_{mn}   + \left(\frac32 - \frac{a_{rs}h^r h^s}{a}\right)\hat{k}  +\frac{2 }{a } h^mh^n \hat{k}_{mn} \\
 = (\lambda -1) \left( \hat{k}+ \frac32 \hat{k} - \frac12 \lambda \hat{k}\right),
\end{align}
which implies
\begin{align}
(\lambda^2 - 6 \lambda + 9) \hat{k} = 0,
\end{align}
which implies for $\hat{k}\neq 0$ that $\lambda = 3$.  If $\hat{k}= 0$, then \eqref{sec} becomes
\begin{align} \label{sum}
 ( \chi^{mn}_{ij}  + \mathcal{M}_{ij}^{mn} ) \hat{k}_{mn}  = \frac12 \left(\lambda^2 -3 \lambda + 4 \right) \hat{k}_{ij}.
\end{align}

We need the eigenvalues for $\mathcal{M}_{ij}^{mn}$. We note that this tensor is not symmetric, since

\begin{align}
  \mathcal M _{ p q r s } -  \mathcal M _{ rspq }   = \frac{2}{a} (a_{rs} h_p h_q - a_{pq} h_r h_s) \neq 0.
\end{align}
However 
\begin{align}\label{S}
S_{ p q r s } = \mathcal M _{ p q r s } - \frac{2}{a} h_p h_q a_{rs},
\end{align}
is symmetric and thus diagonalisable and with real eigenvalues. For trace-free $k^{rs}$ we have from \eqref{S} that
\begin{align}
S_{ p q r s } k^{rs} =  \mathcal M _{ p q r s } k^{rs},
\end{align}
which implies that \eqref{sum} is equivalent to 
\begin{align} \label{replaced}
 ( \chi^{mn}_{ij}  + S_{ij}^{mn} ) \hat{k}_{mn}  = \frac12 \left(\lambda^2 -3 \lambda + 4 \right) \hat{k}_{ij}.
\end{align}

We now obtain the eigenvalues of $S_{ p q r s}$ using the basis vectors $e^i$, $f^i$ defined in \eqref{basisvectors} and $h^i$ .

We have the following eigentensors with corresponding eigenvalues $s_1$-$s_6$:
\begin{align}
(X_1)^{mn} &= h^m h^n,  \quad s_1 = - \frac{4a_{rs} h^r h^s}{a},\\
(X_2)^{mn} &= e^{(m} f^{n)}, \quad  s_2 = \frac{2a_{rs} h^r h^s}{a},\\
(X_3)^{mn} & = e^m e^n - f^m f^n,  \quad s_3 = \frac{2a_{rs} h^r h^s}{a},\\
(X_4)^{mn} & = h^{(m} e^{n)},  \quad s_4 = - \frac{2a_{rs} h^r h^s}{a},\\
(X_5)^{mn} & = h^{(m} f^{n)},  \quad s_5 = - \frac{2a_{rs} h^r h^s}{a},\\
(X_6)^{mn} & = h^2 b^{mn}-h^m h^n, \quad s_6 = - \frac{2a_{rs} h^r h^s}{a}.
\end{align}
As a check compute
\begin{align}
{S_{ij}}^{ij} = -\frac{6a_{rs} h^r h^s}{a} = \sum_{i=1}^6 s_i.
\end{align}

In particular the largest eigenvalue is:
\begin{align}\label{largest}
 s_{max}=s_2 =s_3= \frac{2a_{rs} h^r h^s}{a}.
\end{align}

Returning to \eqref{replaced} and denote the eigenvalues of the tensor $\chi^{mn}_{ij} +S_{ij}^{mn}$ with $\nu$.  Then we have with \eqref{replaced} that
\begin{align}
\lambda^2 -3 \lambda + 4 - 2 \nu = 0,
\end{align}
which implies 
\begin{align}
\lambda = \frac32 \pm \sqrt{-\frac{7}{4} + 2\nu}.
\end{align}
This implies that if $\nu < 1$, then $\mathfrak{Re}{ \lambda} >1$. If $\nu < 2 $, then $\mathfrak{Re}{\lambda} >0$.

Since both $\chi^{mn}_{ij}$ and $S_{ij}^{mn}$ are symmetric $\nu$ will be real and bounded by the largest eigenvalue of $\chi^{mn}_{ij}$ plus the largest eigenvalue of $S_{ij}^{mn}$, which implies using \eqref{mu} and \eqref{largest} that
\begin{align}
\nu \leq \mu + s_{max} =  \frac12 + \frac{3}{a} a_{mn}h^m h^n.
\end{align}
This implies that if  
\begin{align}\label{magneticbound}
a_{mn}h^m h^n < \frac{a}{6},
\end{align}
 then $\mathfrak{Re}{\lambda} >1$ and if $a_{mn}h^m h^n < \frac{a}{2}$, then $\mathfrak{Re}{\lambda} >0$. But we already know that the latter is the case due to \eqref{h2bound}.

We have thus shown that all conditions of Theorem 4 of \cite{LNST} are satisfied. In the case $a_{mn}h^m h^n < \frac{a}{6}$ we have in addition that the real part of the eigenvalues is greater than $1$. 

\medskip

\subsection{The main results of Einstein-Vlasov}

Recall that the unknowns for the Einstein-Vlasov system with a magnetic field are $ 3 \times 3 $ real symmetric matrices, a vector and a distribution function. We are now ready to conclude with the following theorem:

\begin{thm}\label{conformalthm}
Let $ a_0 , b_0 , k_0 , { Z }_0 \in S_2(\bbr^3) $, $h \in \bbr^3 $,  and $ f_0 \geq 0$, a smooth function with compact support in $\mathbb{R}^3\setminus\{ 0 \}$ which is not identically zero,  be initial data of the rescaled Einstein-Vlasov system with a magnetic field with Bianchi I symmetry \eqref{Vlasov}, \eqref{EV1}--\eqref{EV4} satisfying the Fuchsian conditions  \eqref{fuchs1},  \eqref{compVlasov} and the constraints \eqref{mconstraint}, \eqref{Hconstraint}. Then, there exists a time interval $ [ 0 , T ] $ on which the rescaled Einstein-Vlasov system has a unique solution $ a_{ i j } , b^{ i j } , k_{ i j } , {  Z }_{ i j } \in C^0 ( [ 0 , T ] ; S_2(\bbr^3) ) $  and $ f \in C^1 ( [ 0 , T ] ; L^1 ( \bbr^3 ) ) $. If in addition $(a_{ij}h^ih^j)_0 < \frac16 $, then we even have a unique differentiable solution $ a_{ i j } , b^{ i j } , k_{ i j } , {  Z }_{ i j } \in C^1 ( [ 0 , T ] ; S_2(\bbr^3) )$. 
\end{thm}

From this theorem we obtain the existence of the physical metric $ \tilde{ a }_{ i j } $, the rate of change tensor $ \tilde{ k }_{ i j } $ and the physical distribution function $ f $ on a time interval $ ( 0 , T ] $. 

In particular from Theorem \ref{conformalthm} we have that $a_{ij}$ and $k_{ij}$ are $C^0$ functions of $\tau$ from which follows that $ a _ { i j } $ is $ C ^ 1 $ and
$ d a _ { i j } / d \tau$ is $ C ^ 0 $ all as functions of $ \tau $. By Taylor's Theorem we have for small $\tau $ that
\begin{align}
  \label{taylor1} a _ { i j }  &= a _ { 0 i j } + a _ { 1 i j } \tau + o ( \tau ) , \\
  \label{taylor2}  \frac { d a _ { i j } } { d \tau } & = a _ { 1 i j }  + o ( 1 ) ,
\end{align}
where $a _ { 0 i j }$ and $a _ { 1 i j }$ are some constants. Using \eqref{ttau}, \eqref{13a} we have that
\begin{align}\label{phys1}
\tilde{a}_{ij} = \tau ^2 a_{ij} = 2 t  a_{ij}.
\end{align}
As a consequence of  \eqref{taylor1} and \eqref{phys1} we can express $\tilde{a}_{ij}$ in terms of $t$:
\begin{align}
\tilde{a}_{ij} = 2t a _ { 0 i j } + a _ { 1 i j } 2 \sqrt{2} t^{\frac32} +  o ( t^{\frac32} ) .
\end{align}
Now we consider the $t$-derivative of \eqref{phys1} 
\begin{align}
   \frac{d \tilde{a}_{ij}}{dt}  = 2a_{ij} +2t \frac{d\tau}{dt} \frac{d a_{ij}}{d\tau} =  2a_{ij} +(2t)^{\frac{1}{2}}  \frac{d a_{ij}}{d\tau} ,
      \end{align}
which using  \eqref{taylor1}--\eqref{taylor2} becomes
   \begin{align}
   \frac{d \tilde{a}_{ij}}{dt}  =  2a _ { 0 i j } + 3 a _ { 1 i j } (2t)^{\frac{1}{2}} + o ( t^{\frac{1}{2}} ).
      \end{align}
We thus have the asymptotics as $t \rightarrow 0^+$:
\begin{align}
\tilde{a}_{ij} &= 2t a _ { 0 i j } + a _ { 1 i j } 2 \sqrt{2} t^{\frac32} +  o ( t^{\frac32} ), \\
 \tilde{k}_{ij} &=  2a _ { 0 i j } + 3 a _ { 1 i j } (2t)^{\frac{1}{2}} + o ( t^{\frac{1}{2}} ).
 \end{align}
If it turns out that $a_{0ij}h^i h^j < \frac16$ then we have from Theorem \ref{conformalthm} that $a_{ij}$ and $k_{ij}$ are $C^1$ functions of $\tau$ from which follows that $ a _ { i j } $ is $ C ^ 2 $,
$ d a _ { i j } / d \tau$ is $ C ^ 1 $ and $ d ^ 2a _ { i j } / d \tau ^ 2 $ as functions of $ \tau $. By Taylor's Theorem we have for small $\tau $ that
\begin{align}
  a _ { i j }  &= a _ { 0 i j } + a _ { 1 i j } \tau + a _ { 2 i j } \tau ^ 2
  + o ( \tau ^ 2 ) , \\
 \frac { d a _ { i j } } { d \tau } & = a _ { 1 i j } + 2 a _ { 2 i j } \tau + o ( \tau ) ,\\
 \frac { d ^ 2 a _ { i j } } { d \tau ^ 2 }& = 2 a _ { 2 i j } + o ( 1 ) ,
\end{align}
where $a _ { 0 i j }$, $a _ { 1 i j }$ and $a _ { 2 i j } $ are some constants. Doing similar computations as in the previous case we obtain
\begin{align}
\tilde{a}_{ij} & = 2 t a_{0ij} + 2 \sqrt{2} a_{1ij} t^{\frac32} + 4 a_{2ij} t^2+ o(t^2), \\
\tilde{k}_{ij} & = 2  a_{0ij} + 3 \sqrt{2} a_{1ij} t^{\frac12} + 8 a_{2ij} t+ o(t).
\end{align}

However due to \eqref{compVlasov} $a _ { 1 i j } = a _ { 2 i j } = 0$. We thus have established Theorem \ref{phystheorem}.

 \medskip

\section{The Einstein-Boltzmann system with a magnetic field}
For this section we need to change the time coordinate again. Recall that the Boltzmann equation with the chosen scattering cross-section (\ref{13}) becomes (see (36)--(37) of \cite{LNST})
\be\label{23}\frac{\partial f}{\partial \tau}= \tau^{\gamma-2}\int_{{\mathbb{R}}^3}\int_{{\mathbb{S}}^2} \frac{h^{2-\gamma}}{\sqrt{a}(b^{mn}p_mp_n)^{1/2}(b^{rs}q_rq_s)^{1/2}}(f(p')f(q')-f(p)f(q))d\omega d^3q,\ee
which is not regular at $\tau=0$ and where $h$ denotes the unphysical relative momentum which should not be confused with the magnetic field. We replace $\tau$ by $s$ where $ds=\tau^{\gamma-2}d\tau$ so that
\be(\gamma-1)s=\tau^{\gamma-1},\ee
and note that $\gamma-1>0$ by assumption. Later it will be useful to use the notation 

\begin{align}\label{cgamma}
c_{\gamma} = \frac{1}{\gamma-1}.
\end{align}

With the time variable change (\ref{23}) becomes
\be\label{24}
\frac{\partial f}{\partial s}=Q(f,f),\ee
with
\be\label{25}Q(f,f) = \frac{1}{\sqrt{a}}\int_{{\mathbb{R}}^3}\int_{{\mathbb{S}}^2}\frac{h^{2-\gamma}}{p^0q^0}(f(p')f(q')-f(p)f(q))d\omega d^3q,\ee
which is regular at $s=0$.

We need to find the Einstein equations in terms of $s$ rather than $\tau$. We start with another redefinition of the rate-of-change tensor: set
\begin{align}
K_{ij}:=\frac{d}{ds}a_{ij},
\end{align}
 so that
\begin{align}
K_{ij}= \tau^{2-\gamma} k_{ij},
\end{align}
and the first two equations in the system are
\bea
\frac{d}{ds}a_{ij}&=&K_{ij}\label{s1},\\
\frac{d}{ds}b^{ij}&=&-b^{im}K_{mn}b^{nj}\label{s2}.
\eea

We redefine $Z_{ij}$ with the aid of (\ref{19}) as
\be\label{ss3}
\hZ_{ij}=\tau^{2-\gamma}Z_{ij}=\frac{c_{\gamma}}{s} \left(\frac{8\pi}{\sqrt{a}}\int fp_ip_j\frac{d^3p}{(b^{mn}p_mp_n)^{1/2}}
+\frac{2}{a}(a_{ij}a_{mn}-2a_{im}a_{jn})h^mh^n-a_{ij}\right)    ,\ee
where we have used the notation \eqref{cgamma}.
It follows then 
\be\label{s3}\frac{d}{ds}K_{ij}=-\frac{\gamma c_{\gamma}}{s} K_{ij}+ \frac{2 c_{\gamma}}{s} \hZ_{ij}-\frac{c_{\gamma}K}{s} a_{ij}-\frac12KK_{ij}+K_{im}b^{mn}K_{nj}
+2\Lambda \left(\frac{s}{c_{\gamma}}\right)^{4c_\gamma-2}a_{ij},\ee
where $K=b^{ij}K_{ij}$.
The derivative of $\hZ_{ij}$ is as follows:
\bea \label{s4}
s\frac{d}{ds}\hZ_{ij}&=&- \hZ_{ij}- c_{\gamma} K_{ij}+ c_{\gamma} \chi^{pq}_{ij}K_{pq} -c_{\gamma} b^{mn} \Psi_{ij} K_{mn}+\frac{8\pi c_{\gamma} }{\sqrt{a}}\int \frac{\partial f}{\partial s}\frac{p_ip_j}{(b^{mn}p_mp_n)^{1/2}}d^3p\\
\nonumber&&+\frac{2c_{\gamma} }{a} \left[ a_{ij}K_{mn}+K_{ij}a_{mn}-2a_{im}K_{jn}-2K_{im}a_{jn}-\frac{2K}{a}(a_{ij}a_{mn}-2a_{im}a_{jn}) \right]h^mh^n.\
\eea

In matrix form we have
\begin{align}
&\frac{d}{ds}  
\left(
 \begin{matrix}
 a_{ij} \\ b^{ij}
 \end{matrix}
\right) =  \left(
 \begin{matrix}
 K_{ij} \\ -b^{im}b^{jn}K_{mn} 
 \end{matrix}
 \right), \\
&s \frac{d}{ds}  
\left(
 \begin{matrix}
 K_{ij} \\ Z_{ij}
 \end{matrix}
\right) + c_{\gamma} 
 \left(
 \begin{matrix}
 \gamma  \delta^{mn}_{ij}+ 3 \pi_{ij}^{mn} & - 2  \delta^{mn}_{ij}\\ 
 \delta^{mn}_{ij} -  \chi^{mn}_{ij}   + \Pi_{ij}^{mn}   - \mathcal{M}_{ij}^{mn}   & \delta^{mn}_{ij}\end{matrix}
 \right) \
 \left(
 \begin{matrix}
K_{mn} \\ Z_{mn}
 \end{matrix}
\right)
 = s \left(
 \begin{matrix}
G_{ij} \\ 0
 \end{matrix}
\right)+ \left(
 \begin{matrix}
0 \\ H_{ij}
 \end{matrix}
\right),
\end{align}
with
\begin{align}
G_{ij} & =  -\frac12KK_{ij}+K_{im}b^{mn}K_{nj}+2\Lambda \left(\frac{s}{c_{\gamma}}\right)^{4c_\gamma-2}a_{ij},\\
H_{ij} & =\frac{8\pi c_{\gamma} }{\sqrt{a}}\int \frac{\partial f}{\partial s}\frac{p_ip_j}{(b^{mn}p_mp_n)^{1/2}}d^3p.
\end{align}

We again have it in the form of Theorem 4 of \cite{LNST} with
\begin{align}
x = 
\left(
\begin{array}{c}
a_{ij} \\
b^{ij}
\end{array}
\right), \quad 
y = 
\left(
\begin{array}{c}
K_{ij} \\
\hat{Z}_{ij}
\end{array}
\right), \quad  F= \left(
 \begin{matrix}
 K_{ij} \\ -b^{im}b^{jn}K_{mn} 
 \end{matrix}
 \right), \quad G =
\left(
\begin{array}{c}
G_{ij} \\
0
\end{array}
\right), \quad H =\left(
\begin{array}{c}
0 \\
H_{ij}
\end{array}
\right),
\end{align}
and 
\begin{align}\label{Nb}
N=  c_{\gamma}  \left(
 \begin{matrix}
 \gamma \delta^{mn}_{ij} + 3 \pi_{ij}^{mn} & - 2 \delta^{mn}_{ij} \\ 
 \delta^{mn}_{ij}-  \chi^{mn}_{ij}   + \Pi_{ij}^{mn}   - \mathcal{M}_{ij}^{mn}   &  \delta^{mn}_{ij} \end{matrix}
 \right).
 \end{align}

The Fuchsian condition \eqref{fuchs1} considered in the Vlasov case will still hold and the compatibility condition is now

\begin{align}\label{compBoltz}
 c_{\gamma}  \left(
 \begin{matrix}
 \gamma \delta^{mn}_{ij} + 3 \pi_{ij}^{mn} & - 2 \delta^{mn}_{ij} \\ 
 \delta^{mn}_{ij}-  \chi^{mn}_{ij}   + \Pi_{ij}^{mn}   - \mathcal{M}_{ij}^{mn}   &  \delta^{mn}_{ij} \end{matrix}
 \right)\left(
\begin{array}{c}
K_{ij} \\
\hat{Z}_{ij}
\end{array}
\right) =  \left(
\begin{array}{c}
0 \\
H_{ij}
\end{array}
\right).
 \end{align}
 
 Following the argument in the Vlasov case, having in mind \cite{LNST} and that we will show the positivity of the real part of the eigenvalues of the new $N$, given $(f,h_i)$ there will be an $a_{ij}$, and these data will give us unique $K_{ij}$, $\hat{Z}_{ij}$ due to \eqref{compBoltz} and since $N$ is invertible. We can conclude:

\begin{prop}\label{propBoltz}
 Let $f_0 \geq 0$ be a smooth function with compact support in $\mathbb{R}^3\setminus\{ 0 \}$. Suppose that
$f_0$ is not identically zero. Then, there exist unique $ 3 \times 3$ symmetric matrices $a_0$, $b_0$, $K_0$ and $\hat{Z}_0$ and a vector $h$
satisfying the Fuchsian conditions \eqref{fuchs1} and  \eqref{compBoltz}. 
 \end{prop}

The differentiability conditions were shown in \cite{LNST} and we already have seen that in the Vlasov case they are satisfied for the magnetic term. We now proceed to compute the eigenvalues of the new $N$ which we again call $\lambda$. The eigentensors of $N$ satisfy

\begin{align}
\gamma K_{ij}+ 3 \pi_{ij}^{mn} K_{mn}  - 2 \hat{Z}_{ij} = \frac{\lambda}{c_\gamma}  K_{ij}, \\
K_{ij}-  \chi^{mn}_{ij}K_{mn}   + \Pi_{ij}^{mn} K_{mn}   - \mathcal{M}_{ij}^{mn}  K_{mn} + \hat{Z}_{ij} =  \frac{\lambda}{c_\gamma} \hat{Z}_{ij}.
\end{align}
From the first equation we have
\begin{align}
\hat{Z}_{ij} = \frac{\gamma}{2} K_{ij}+ \frac32 \pi_{ij}^{mn} K_{mn} - \frac{\lambda}{2c_\gamma} K_{ij},
\end{align}
so that the second equation becomes
\begin{align}
K_{ij}-  \chi^{mn}_{ij}K_{mn}   + \Pi_{ij}^{mn} K_{mn}   - \mathcal{M}_{ij}^{mn}  K_{mn}  = \left (\frac{\lambda}{c_{\gamma}} -1 \right)  \left(  \frac{\gamma}{2} K_{ij}+ \frac32 \pi_{ij}^{mn} K_{mn} - \frac{\lambda}{2c_\gamma} K_{ij}\right),
\end{align}
which using the definitions of $\chi^{mn}_{ij}$, $\Pi_{ij}^{mn}$ and $\pi_{ij}^{mn}$ turns into
\begin{align}\label{second}
K_{ij}-  b^{mr}b^{ns}\Psi_{ijrs}  K_{mn}   + \Psi_{ij} \hat{k}   - \mathcal{M}_{ij}^{mn}  K_{mn}  = \left (\frac{\lambda}{c_{\gamma}} -1\right) \left( \frac{\gamma}{2} K_{ij}+ \frac12 a_{ij} K - \frac12 \frac{\lambda}{c_{\gamma}} K_{ij}\right).
\end{align}

Contract with $b^{ij}$ 
\begin{align}
2K   = \left (\frac{\lambda}{c_{\gamma}} -1 \right) K  \left( \frac{\gamma}{2} + \frac32  - \frac12 \frac{\lambda}{c_{\gamma}}  \right),
\end{align}
which implies
\begin{align}
\left[ \left(\frac{\lambda}{c_{\gamma}} \right )^2 - (4+\gamma) \frac{\lambda}{c_{\gamma}} + \gamma+7 \right] K = 0,
\end{align}
which implies for $K\neq 0$ that 
\begin{align}
\lambda = c_{\gamma} \left( 2+\frac12\gamma \pm \sqrt{ -3 +  \gamma + \frac14 \gamma^2 } \right).
\end{align}
We have that $\gamma$ is between $1$ and $2$ and we have that $-3 +  \gamma + \frac14 \gamma^2 $ is zero for $\gamma=2$. Thus the real part of $\lambda$ is always bigger than
\begin{align}
(2+\frac12\gamma)  c_{\gamma} = \frac{4+ \gamma}{2 (\gamma + 1)},
\end{align}
which is always greater than one.
  If $K= 0$, then \eqref{second} becomes
\begin{align}
( \chi^{mn}_{ij}  + \mathcal{M}_{ij}^{mn} ) K_{mn}  = \frac12 \left(\frac{\lambda^2}{c_\gamma^2} -(\gamma+1) \frac{ \lambda}{c_{\gamma}} + \gamma+2 \right) K_{ij},
\end{align}
and we obtain
\begin{align}
\lambda  = c_{\gamma} \left( \frac{\gamma+1}{2} \pm \frac12 \sqrt{ \gamma^2 -2 \gamma - 7 + 8 \nu} \right).
\end{align}
We have that
\begin{align}
c_{\gamma}  \frac{\gamma+1}{2} \geq \frac32.
\end{align}
Now using the bound on $\nu$ and the bound on \eqref{h2bound} we obtain
\begin{align}
 \gamma^2 -2 \gamma - 7 + 8 \nu \leq \gamma^2 -2 \gamma - 3 + \frac{24}{a} a_{mn}h^m h^n \leq  - 3 + \frac{24}{a} a_{mn}h^m h^n.
\end{align}
We have eigenvalues with a positive real part if  
\begin{align}
- 3 + \frac{24}{a} a_{mn}h^m h^n < 9 \iff a_{mn}h^m h^n < \frac a2,
\end{align}
which is always the case and we have eigenvalues with real part greater than one if
\begin{align}
- 3 + \frac{24}{a} a_{mn}h^m h^n < 1  \iff a_{mn}h^m h^n < \frac a6,
\end{align}
so we have for the Boltzmann case the same conditions as in the Vlasov case.

Since we have now the existence of solutions to the Einstein part, we can apply \cite{LNST} to couple it to the Boltzmann part. Before establishing the theorem we need for the distribution function  as in \cite{LNST} the following weighted $ L^p $-spaces. Let $ L^1_r ( \bbr^3 ) $ and $ L^\infty_\eta ( \bbr^3 ) $ denote the spaces of functions equipped with the following norms:
\begin{align}
& \| f \|_{L^1_r} = \int_{\bbr^3} | f(p) | ( p^0)^r \, d^3p,\qquad p^0 = \sqrt{b^{ij} p_i p_j}, \\
& \| f \|_{ L^\infty_\eta} = \sup_{ p \in \bbr^3 } | w_\eta f ( p ) | , \qquad w_\eta = p^0 \exp ( s^{ - 1 }_\eta p^0 ) , \qquad s_\eta = ( s + \eta^2 )^\eta , \qquad \eta > 0 .
\end{align}

We are now ready to establish the following theorem:

\begin{thm}\label{conformalpropBoltz}
Let $ a_0 , b_0 , K_0 , { \hat Z }_0 \in S_2(\bbr^3) $, $h \in \bbr^3$   and $ 0 \leq f_0 \in L^1 ( \bbr^3 ) $ be initial data of the rescaled Einstein-Boltzmann system with a magnetic field \eqref{24}--\eqref{25},  \eqref{s1}--\eqref{s4} with Bianchi I symmetry, satisfying the Fuchsian conditions \eqref{fuchs1}, \eqref{compBoltz} and the constraints \eqref{mconstraint},\eqref{Hconstraint}. There exist small positive $\delta$ and $\eta$ such that if
\begin{align}
f_0 \in L^1_1( \bbr^3 ) \cap L^1_{ - 2 - \delta / 2 } ( \bbr^3 ) \cap L^\infty_\eta ( \bbr^3) , \qquad \frac{ \partial f_0 }{ \partial p } \in L^1_1 ( \bbr^3 ) \cap L^1_{ - 1 - \delta / 2 } ( \bbr^3 ) ,
\end{align}
then, there exists a time interval $ [ 0 , T ] $ on which the rescaled Einstein-Boltzmann system with a magnetic field has a unique solution $ a_{ i j } , b^{ i j } , K_{ i j } , { \hat Z }_{ i j } \in C^0 ( [ 0 , T ] ; S_2(\bbr^3) ) $ and $ 0 \leq f \in C^1 ( [ 0 , T ] ; L^1 ( \bbr^3 ) ) $.   If $(a_{ij}h^ih^j)_0 < \frac16 $ then we even have a unique differentiable solution such that  $ a_{ i j } , b^{ i j } , K_{ i j } , { \hat Z }_{ i j } \in C^1 ( [ 0 , T ] ; S_2(\bbr^3) ) $.
\end{thm}

We can now translate this to the physical version using the results of \cite{LNST} and obtain Theorem \ref{phystheoremboltz}.

\section*{Conflicts of interest/Competing interests}
The authors acknowledge the hospitality from the Oberwolfach Research Institute for Mathematics where during the research stay under the Oberwolfach Research Fellow
(OWRF) program with reference number 2427q the present work was finished in an excellent working environment. This work was supported by the National Research Foundation of Korea (NRF) grant funded by the Korea government (MSIT) (No. RS-2024-00451692). Otherwise, the authors have no conflicts of interest to declare that are relevant to the content of this article.

\end{document}